# Artificial confocal microscopy for deep label-free imaging

Xi Chen[1], Mikhail E. Kandel[1,2], Shenghua He[3], Chenfei Hu[1,4], Young Jae Lee[1,5], Kathryn Sullivan[6], Gregory Tracy[7], Hee Jung Chung[5,7], Hyun Joon Kong[1,6,8,9], Mark Anastasio[1,6], Gabriel Popescu[1,4,6,9]

[1]Beckman Institute for Advanced Science and Technology, University of Illinois at Urbana-Champaign, Urbana, IL, USA.

[2]Currently with Groq, 400 Castro St., Suite 600, Mountain View, CA 94041, USA

[3]Department of Computer Science & Engineering, Washington University in St. Louis, St. Louis, Missouri, 63130, USA

[4]Department of Electrical and Computer Engineering, University of Illinois at Urbana-Champaign, Urbana, IL, USA.

[5]Neuroscience Program, University of Illinois at Urbana-Champaign, Urbana, IL, USA.

[6]Department of Bioengineering, University of Illinois at Urbana-Champaign, Urbana, IL, USA.

[7]Department of Molecular and Integrative Physiology, University of Illinois at Urbana-Champaign, Urbana, IL, USA

[8]Chemical and Biomolecular Engineering, University of Illinois at Urbana-Champaign, Urbana, IL, USA.

[9]Carl Woese Institute for Genomic Biology, University of Illinois at Urbana-Champaign, Urbana, IL, USA.

The official email addresses of all authors:

Xi Chen: xic@illinois.edu

Mikhail E. Kandel: kandel3@illinois.edu

Shenghua He: shenghuahe@wustl.edu

Chenfei Hu: chenfei3@illinois.edu

Young Jae Lee: lee134@illinois.edu

Kathryn Sullivan: kmsllvn2@illinois.edu

Gregory Tracy: gtracy4@illinois.edu

Hee Jung Chung: chunghj@illinois.edu

Hyun Joon Kong: hjkong06@illinois.edu

Mark Anastasio: maa@illinois.edu

Gabriel Popescu: gpopescu@illinois.edu

Corresponding author:

Gabriel Popescu, 4055 Beckman Institute, 405 North Mathews Ave, Urbana, Illinois 61801, (217) 333-4840, gpopescu@illinois.edu

**Abstract:**

Widefield microscopy methods applied to optically thick specimens are faced with reduced contrast due to "spatial crosstalk", in which the signal at each point in the field of view is the result of a superposition from neighboring points that are simultaneously illuminated. In 1955, Marvin Minsky proposed confocal microscopy as a solution to this problem. Today, laser scanning confocal fluorescence microscopy is broadly used due to its high depth resolution and sensitivity, which come at the price of photobleaching, chemical, and photo-toxicity. Here, we present artificial confocal microscopy (ACM) to achieve confocal-level depth sectioning, sensitivity, and chemical specificity, on unlabeled specimens, nondestructively. Thus, we augmented a commercial laser scanning confocal instrument with a quantitative phase imaging module, which provides optical pathlength maps of the specimen on the same field of view as the fluorescence channel. Using pairs of phase and fluorescence images, we trained a convolution neural network to translate the former into the latter. The training to infer a new tag is very practical as the input and ground truth data are intrinsically registered and the data acquisition is automated. Remarkably, the ACM images present significantly stronger depth sectioning than the input (phase) images, enabling us to recover confocal-like tomographic volumes of microspheres, hippocampal neurons in culture, and 3D liver cancer spheroids. By training on nucleus-specific tags, ACM allows for segmenting individual nuclei within dense spheroids for both cell counting and volume measurements. Furthermore, taking the estimated fluorescence volumes, as annotation for the phase data, we extracted dry mass information for individual nuclei. Finally, our results indicate that the network learning can be transferred between spheroids suspended in different media. In sum, ACM can provide quantitative, dynamic data, nondestructively, from both thin and thick samples, while chemical specificity is recovered computationally.

**Introduction**

Three-dimensional (3D) cellular systems have been increasingly adopted over 2D cell monolayers to study disease mechanisms[1] and discover drug therapeutics[2], as they more accurately recapitulate the *in vivo* cellular function and development of extracellular matrices[3]. 3D cellular structures, including cellular clusters such as organoids and spheroids, have found use in a wide range of applications such as tissue engineering[4], high-throughput toxicology[5], and personalized medicine[6]. A particularly exciting direction of research is engineering multicellular living systems[7-9]. These fields of current scientific interest bring along the urgent need for new methods of investigation to inform about cellular viability and cell cluster proliferation. Ideally, such techniques would provide quantitative data with subcellular resolution at arbitrary depths in the cellular system and dynamic information rendered over broad time scales. Importantly, these assays would be completely nondestructive, i.e., would report on the cell cluster without interfering with its viability and function.

Due to the visible light wavelengths being comparable with the subcellular structures, optical methods of investigation are well suited for meeting these requirements. However, significant challenges exist for the existing optical microscopy techniques when applied to increasingly thick samples. Typical spheroids, ranging in size from hundreds of microns to millimeters, are significantly larger than the scattering mean free path associated with the light wave propagation, generate strong multiple scattering, and, thus, form optically turbid aggregates that are difficult to analyze at a cellular level[10,11]. As a result, high-throughput investigations often are limited to extracting coarse parameters, such as spheroid diameters, at low-magnification[12].

In 1955, in his pursuit to image 3D biological neural networks and mimic their behavior computationally, Minsky was faced with the challenge of suppressing multiple scattering, which was particularly severe for the widefield instruments available at the time[13]. In Minsky's own words, "One day it occurred to me that the way to avoid all that scattered light was to never allow any unnecessary light to enter in the first place. An ideal microscope would examine each point of the specimen and measure the amount of light scattered or absorbed by that point."[13] This first implementation of the confocal scanning microscope was established in a transmission geometry, requiring sample translation. Of course, today's modern confocal instruments take advantage of bright laser sources, use beam scanning, and are most often used in a reflection geometry, paired with fluorescence contrast[14]. In time, many other advanced laser scanning techniques have been developed for fluorescence microscopy[15]. Nevertheless, fluorescence imaging is subject to several limitations. Absorption of the excitation light may cause the fluorophore to photobleach, which limits the time interval over which continuous imaging can be performed[16]. The excitation light is typically toxic to cells, a phenomenon referred to as phototoxicity, while the exogenous fluorophores themselves can induce chemical toxicity[17]. While the advancement of the green fluorescent protein technology significantly improves the viability of the specimen under investigation, concerns regarding phototoxicity, photobleaching, and functional integrity of the cells upon genetic engineering still remain[18]. Overcoming these limitations becomes extremely challenging when imaging thick objects over an extended period of time and, for that reason, confocal microscopy is most regularly used on fixed specimens [19,20].

Recently, label-free nonlinear imaging techniques have been used to extract structural information with subcellular resolution[21]. The limitations associated with fluorescence imaging have motivated the development of interferometric label-free imaging modalities, which rely on intrinsic, scattering contrast. Optical coherence tomography (OCT) was reported to detect and count aqueous cells in the anterior chamber of a rodent model of eye inflammation[22]. OCT was also used to measure necrotic regions and volumetrically quantify tumor spheroids[23]. Several phase-sensitive methods developed in a confocal modality have been recently developed, but their application to thick structures has been mostly unexplored[24-27]. Quantitative phase imaging (QPI)[28] has emerged recently as a potentially valuable label-free approach, which, due to its high resolution and sensitivity, has found a broad range of new applications[29]. While most of QPI's applications involve thin specimens (cell monolayers, thin tissue slices)[30-32], gradient light interference microscopy (GLIM)[33,34] has been developed to suppress multiple scattering via white light, phase-shifting interferometry, which allowed for imaging and analyzing quantitatively opaque structures, such as spheroids and embryos. However, as a widefield technique, GLIM has limited axial resolution and suffers from the spatial crosstalk described early-on by Minsky, which mixes diffraction contributions by neighboring points from within the specimen. As a result, the accurate discrimination of cellular boundaries deep within a spheroid remains challenging.

In this paper, we report artificial confocal microscopy (ACM), a laser scanning QPI system combined with deep learning algorithms, which renders "synthetic" fluorescence confocal images from unlabeled specimens. Our work builds on recent breakthrough developments that apply deep learning tools to enhance optical imaging[35-42]. Interestingly, our work combines the two fields pioneered by Minsky in the 1950s: confocal microscopy and artificial intelligence (AI), as follows. First, we developed a laser-scanning QPI system, which is implemented as an upgrade module onto an existing laser scanning confocal microscope (LSM 900, Airyscan 2, Zeiss). We validated the boost in sensitivity and axial resolution of the new system by using standard samples and rigorous comparison with the widefield counterpart. Second, we derived a theoretical model based on the first-order Born approximation, which yields an analytic solution for the spatial frequency coverage of the laser scanning QPI system. These results were validated using experiments to measure the transfer function of the instrument. Third, we trained an artificial neural network on pairs of laser scanning QPI and fluorescence confocal images from the same field of view. Because the QPI module is attached to the same optical path, generating the training data is straightforward and automated, as the fields of view are intrinsically registered. Fourth, we applied the inference of the "computational" neural network to monolayers of "biological" neural networks and found that the resulting 3D images mimic very well those of the ground truth from the confocal fluorescence images. Using these ACM images, we created binary masks for the contour of the cell and applied them back into the QPI (input) data. Our results show that the measurements of cell volume and dry mass of ACM vs. confocal agree very well. Fifth, we used the ACM images to perform nuclear segmentation and, thus, cell counting, within hepatocyte spheroids. We also showed that the training performed on spheroids suspended in PBS transfers well to specimens suspended in hydrogel, which promises broad applications in tissue engineering.

**Results**

The ACM imaging system consists of an existing confocal microscope augmented by a laser scanning GLIM system (LS-GLIM). Figure 1a illustrates the ACM setup, which has three main modules: the laser-scanning confocal microscope (LSM 900, Zeiss), the differential interference contrast (DIC) microscope, and the LS-GLIM module. The LS-GLIM assembly shares the laser source from confocal microscopy (see Methods). The two sheared beams that form the DIC image have their relative phase shift controlled by the liquid crystal variable retarder (LCVR). The LCVR was carefully calibrated to produce accurate phase shifts, as described in SI Note 1. For each $\pi/2$ phase shift, the transmitted light photomultiplier tube (PMT) records the resulting interferogram, as shown in Fig. 1b. The quantitative phase images are generated by the phase-retrieval reconstruction and Hilbert integration algorithms described in the GLIM opperation[34]. By sharing the same illumination path, the imaging system registers QPI z-stack images and pairs them with confocal fluorescence frames from the same field of view, which serve, respectively, as input and ground truth data for the deep learning algorithm (Figs. 1b-c). Due to the laser-scanning illumination and PMT detection, the noise level is reduced by a factor of 5 compared to the full-field method (See SI Note 2), thus, the spatial sensitivity of the phase images is improved. The ground truth data, i.e., confocal fluorescence images, provide specificity with high axial resolution and signal-to-noise ratio (SNR). Our goal is to use deep learning to infer the fluorescence confocal images from the LS-GLIM input data and, thus, replicate the confocal advantages on unlabeled specimens.

Multichannel EfficientNet-based U-Nets (E-U-Nets) were trained to translate the 3D phase image stack to the corresponding 3D fluorescent image stack. An E-U-Net comprises a standard U-Net where the encoder is replaced with an EfficientNet[43] (Fig. 2a). The multichannel inputs of an E-U-Net are 3 neighboring quantitative phase images along the z-axis, and the output is the corresponding central fluorescent image slice (see Methods). We chose this three-frame set as input to account for the fact that the axial spread in LS-GLIM data is significantly more pronounced than in the confocal fluorescence data, primarily because the input image is obtained in a transmission geometry without a pinhole, while the output is in reflection with a pinhole or Airyscan detector array. Thus, the neural network "learns" the spread mechanism from the three adjacent images and reverses it to produce a sharp ACM frame.

In SI Note 3, we present a full description of the 3D image formation in LS-GLIM, which starts with the inhomogeneous wave equation and considers scattering under the Born approximation. The expression for the signal collected at the detector has a particularly simple and physical intuitive form,

$$s(\boldsymbol{\rho}) \propto \chi(\boldsymbol{\rho}) \circledv \left[ U_d(\boldsymbol{\rho}) U_i^*(\boldsymbol{\rho}) \right], \qquad [1]$$

where $\chi$ is the scattering potential of the specimen, $U_d$ and $U_i$ are the detection and illumination functions, defined as the Fourier transforms of their respective pupil functions, * stands for complex conjugation, and $\circledv$ denotes the 3D convolution in the spatial domain, $\boldsymbol{\rho}$. Thus, the point spread function is given by the product $U_d(\boldsymbol{\rho}) U_i^*(\boldsymbol{\rho})$, i.e., it improves with both a tighter illumination focus and a broader detection pupil. These theoretical predictions are comparable with the experimental measurements of the LS-GLIM transfer function for various detection NAs (see Supplementary Fig. S4).

Figure 2b compares images of a 2 $\mu m$ microbead under widefield GLIM, LS-GLIM, confocal fluorescence microscopy, and the network inference, i.e., the ACM image. Interestingly, the resulting ACM image is characterized by significantly less axial blur than the LS-GLIM input. As described in the SI Note 2, the sensitivity of LS-GLIM is superior to its widefield counterpart, due to the absence of spatial crosstalk and higher sensitivity photon multiplier detector. However, due to the transmission geometry, they are both inferior to the reflection confocal images in terms of axial sectioning. In contrast, the corresponding network inferences, i.e., the ACM images show significantly improved axial resolution and sectioning. The three adjacent LS-GLIM frames used as network input (see methods for details) contain information about the field Laplacian along z, which governs the inhomogeneous wave equation (Supplemental Note 3) and may explain why this network architecture can produce adequate results in terms of 3D reconstructions.

Next, we applied ACM to imaging neural cultures. We used two common stains to tag the Tau and MAP2 proteins[44] (see Methods) the ratios of which are a popular model for differentiating the long axon from smaller dendrites. The confocal fluorescence images from the two channels represent the ground truth and, as before, the corresponding LS-GLIM images were the input data. The results are summarized in Figs. 3a-l. Our results indicate that the overall 3D renderings of the ground truth and their inferences match very well. Occasionally, we found some discrepancies in the dendrites, which translates into lower correlation values (see SI Note 4). In Figs. 3f, l, the white arrow points to the axon of the neuron. The PSNR and Pearson correlation (between estimated and actual fluorescence) are summarized in Supplementary Note 4. Interestingly, ACM images reduced the pixel-level noise and confocal stripe artifacts present in the training data. The ACM data allows us to delineate individual cells accurately and measure their volumes. Supplemental Video 1 illustrates this performance on live neurons that have never been labeled. Visually, it is evident that the ACM provides a much sharper decay of the out-of-focus light, i.e., greater depth sectioning, than the original LS-GLIM. Supplemental Video 2 illustrates the time-lapse performance of ACM on unlabeled, dynamic neurons. Of course, the ACM images do not suffer from bleaching or toxicity, while they maintain chemical specificity through computation. As a result, ACM is ideal for studying live cellular systems non-destructively, over large periods of time.

From the ACM images, we computed binary masks corresponding to the cell contours, which were applied back to the input QPI maps to retrieve individual cell dry mass values. From the cell volume and mass, we also extracted the dry mass density for each cell. The neuron volume and dry mass density measurements are illustrated in Figs. 3m-t. The training data contained 20 z-stacks of neurons at 10 days in vitro (DIV 10). Figure 3s shows the box plots of the volume measurements from the ground truth and ACM. The values plotted represent the cell volume averaged per field of view. Our results indicate that the volume measurements are well-matched with the ground truth, i.e., there is no significant difference between the two distributions ($p \gg 0.05$).

To demonstrate ACM's ability to delineate cellular structures inside turbid spheroids, we imaged hepatocyte spheroids (HepG2) suspended in phosphate-buffered saline (PBS) and generated computational stains associated the deoxyribonucleic acid (DNA) and ribonucleic acid (RNA) (Fig. 4a). The RNA is localized within the nucleus, with a high concentration in the nucleolus (Fig. 4b). The study of RNA is of high current interest, not only because it plays a crucial role in catalyzing cellular processes, but also because it can be used by various viruses to encode their genetic information[45]. The two ground truth stains, 7-aminoactinomycin D (7-ADD) and SYTO RNASelect Green (see Methods), and their associated inferences, enable us to generate semantic segmentations and annotate the spheroid into "nuclei" and "nucleoli", respectively. The entire "spheroid" represents our third class and is obtained as the non-background regions in the LS-GLIM data. As shown in Fig. 4c and SI Note 4, the actual and imputed fluorescent maps show good agreement. As detailed in Methods, we apply a threshold on the phase image to generate 3D semantic segmentation maps, which we use to measure the dimension of the spheroid. The intersection of the RNA and DNA labels provides the annotation for the nucleoli. Our results show that, across the 20 spheroids studied in this work, the total nuclear mass is proportional to the spheroid mass. This dependence is shown in Fig. 4d, where the slope of the linear regression (0.42) indicates that about 42% of the spheroid mass is contributed by the nuclei.

Automatic instance segmentation of cells inside spheroids was performed by 3D marker-controlled watershed on the estimated DNA signal. The markers were determined through a 2D

Hough voting on each slice in the z-stack basis (Fig. 5). The result of the Hough voting is a volume with a unique marker on the spheroid, which resembles a column tracking the center of the nucleus through the focus (see Methods for details). The result of the watershed is a 3D volume with a unique label for each nucleus within the spheroid, which enables the calculation of parameters on individual cells. In order to compare the mass, volume, and mass density distributions, we computed the relative spread, $\sigma/\mu$, with $\sigma$ the standard deviation and $\mu$ the mean associated with the best Gaussian fit. Our data indicate that nuclear density (Fig. 5d, $\sigma/\mu = 0.2$) has a significantly narrower distribution compared to the distributions of nuclear mass (Fig. 5b, $\sigma/\mu = 0.9$) and volume (Fig. 5c, $\sigma/\mu = 0.8$). These observations indicate that the dry mass density is a much more uniform parameter across different cells. Given the broad distribution of volumes and masses, this result shows that a change in volume is accompanied by an almost linear change in mass.

Since spheroids are often grown in specialized environments, we asked the question of whether the training of our neural network can be transferred from PBS to hydrogel suspensions. Because of the much firmer scaffold that the hydrogel provides, the respective spheroids approach a more spherical shape (Fig. 6). The ACM and ground truth tomography of the 7-ADD tag are shown in Fig. 6. The Pearson correlation and PSNR are presented in SI Note 4. The binary map for each z-position is generated from the inferences of the network through thresholding. Next, the 3D dry mass tomography is obtained with the binary maps and the input phase images as described earlier (see Figs. 6a-h and also Methods). Using the volume and dry mass information, we obtained the 3D dry mass density, as shown in Figs. 6i-l. These results show that the prediction accuracy of our network is not degraded significantly when performing transfer learning from one surrounding medium to another.

**Discussion:**

It is rather remarkable that the principles of AI and confocal microscopy were both formulated in the mid-1950s, with Marvin Minsky standing at the origin of both. Since then, the two technologies have taken on independent trajectories, with confocal leading to an entirely new class of scanning imaging modalities and AI giving rise to a variety of applications, from digital assistants to autonomous vehicles. Furthermore, in the past several years, it has become apparent that AI algorithms are valuable tools in extracting knowledge from optical images. As such, the two fields are intersecting again, and this combination appears to hold exciting prospects for biomedicine.

We developed artificial confocal microscopy to combine the benefits of non-destructive imaging from QPI with the depth sectioning and chemical specificity associated with confocal fluorescence microscopy. Augmenting an existing laser scanning microscope with a QPI module (LS-GLIM) we can easily collect pairs of registered images from the phase (input data) and fluorescence (ground truth) channels. As expected, the transmission quantitative phase image exhibits a much stronger elongation along the z-axis, as the scattering wavevector (or momentum transfer) has a much shorter z-component than in the reflection geometry. These pairs of images are used to train a neural network (Efficient U-Net) to perform image-to-image translation from the LS-GLIM to the confocal fluorescence signal. The final ACM image presents the characteristics of the confocal image, with sharp axial sectioning and chemical specificity (see Figs. 2-3). Applying ACM to unlabeled cells allows us to nondestructively translate the confocal microscopy features to dynamic imaging (see Supplemental Videos 1-2). The image formation in LS-GLIM was described here for the first time by assuming a weakly scattering model. The theoretical model appears to be reasonable, as indicated by the comparison with experimental data on the system's transfer function.

By overcoming the spatial crosstalk limitations associated with widefield methods, ACM has the potential to provide new data for studying turbid cellular systems. Measuring quantitatively functional parameters from organoids and spheroids can be useful in a variety of applications of biological and clinical relevance. Using the artificial fluorescence images generated by the neural network, we segmented individual nuclei within the 3D structures, which can be used not just for cell counting but also for computing individual nuclear volumes. Furthermore, by creating annotation from the ACM images and applying them back to the input phase images, we extracted dry mass information from individual nuclei, independently from the nuclear volume. Our results indicate that, on average, 42% of the spheroid mass is contained in the nuclei. We also found that the nuclear dry mass density distribution is significantly narrower than the volume and mass counterparts.

Finally, we demonstrated that the network training can be transferred between spheroids suspended in different media. Thus, we found that ACM performs well on spheroids suspended in hydrogel with no additional training, which provides versatility to our technique. We anticipate that ACM can be potentially adopted at a broad scale because the LS-GLIM module can be readily added to any existing laser confocal system, while the data for training can be acquired with ease. ACM provides complementary information to that from other laser-scanning techniques, as the acquisition is not limited by photobleaching and toxicity, while the axial resolution is maintained at confocal levels.

**Methods**

*ACM system*

The experimental set-up for ACM is a multichannel imaging system, which consists of confocal microscopy (LSM 900, Zeiss) and laser-scanning gradient light interference microscopy (LS-GLIM). The LS-GLIM module upgrades a laser scanning confocal microscope outfitted with DIC optics by providing phase-shifting assembly capability (Fig. 1(a)). The laser-scanning interference microscopy shares the same two-laser lines (488nm, 561nm) of the confocal microscope. The laser source from the confocal microscopy goes up through the matched DIC prism and objective (63×, 40×) and then is scattered by the sample. After the sample, the light is collected by the condenser of the DIC microscope. Next, the light travels through the phase-shifting assembly, which consists of a liquid crystal variable retarder (LCVR, Thorlabs) followed by a linear polarizer. In order to allow the liquid crystal to modulate the phase shift between the two orthogonal polarizations, we removed the analyzer that normally sits inside the condenser. The stabilization time of the LCVR is approximately 70 ms. Four intensity frames are recorded by the photomultiplier tube (PMT, Zeiss) corresponding to each $\pi/2$ phase shift, as shown in Fig. 1(b). The acquisition time of each frame is approximately the same as for a confocal fluorescence image, which depends on the dwell time and pixel numbers set for the image acquisition. The dwell time for all the images was chosen to be 1.2 μs, such that for an image with 1744×1744 pixels, the acquisition time is ~ 3.7s. The quantitative phase images are generated in real-time by the phase-retrieval reconstruction algorithm and Hilbert transform algorithm[34]. The system registers pairs of z-stack images from both the confocal fluorescence and quantitative phase, which serve, respectively, as ground truth and input images for machine learning (Fig. 1(b)). The z-sampling was chosen to be 0.2 $\mu m$, 0.2 $\mu m$, and 1 $\mu m$ for microbeads, neurons, and spheroids, respectively. The x-y sampling was 0.09 $\mu m$ for all the data presented in this paper.

**Network training:**

Multichannel EfficientNet-based U-Nets (E-U-Nets) were trained with paired phase and fluorescent images. The input channels of an E-U-Net are three neighboring phase slices, and the output is the corresponding central fluorescent slice. This network design allows an E-U-Net to use information from phase images acquired at multiple neighboring imaging planes to better predict the fluorescent image.

The network architecture of a multichannel E-U-Net is shown in Fig. 2a and Fig. S6. It represents a modification of a standard U-Net where the encoder is replaced with an EfficientNet[43]. The EfficientNet generally has a powerful capacity of feature extraction but is relatively small in network size. Training an E-U-Net from scratch can be challenging when the number of paired phase and fluorescent images is limited. To mitigate this challenge, a transfer learning strategy was used in the E-U-Net training. Specifically, the weights of the EfficientNet encoder were initialized with weights pre-trained on an ImageNet dataset[46] for an image classification task. The ImageNet is a benchmark image set that contains millions of labeled nature images.

In this study, a neuron dataset, a spheroid cell dataset, and a bead dataset were used for training, validating, and testing the E-U-Nets. The neuron dataset contained 22 image stacks that each contained 300 neuron phase images of size 1744 × 1744 pixels and their related two-channel fluorescent images, which correspond to fluorescent signals from Tau and MAP2 proteins, respectively. The spheroid cell dataset contained 21 stacks that each contained 100 spheroid cell phase images of size 1744×1744 pixels and the related two-channel fluorescent images, which correspond to fluorescent signals from DNA and RNA, respectively. The bead dataset contained 18 image stacks that each contained 250 bead phase images of size 128×128 pixels and the associated fluorescent images. To facilitate network training, the pixel values in each fluorescent image stack were scaled to a range of [0, 255.0]. This was accomplished as: $x_o = 255.0 \times \frac{x_i - x_{0.01\%}}{x_{99.99\%} - x_{0.01\%}}$, where $x_{0.01\%}$ and $x_{99.99\%}$ represent the 0.01%-th and 99.99%-th values among all the pixel values in the image stack after they were sorted in non-decreasing order; $x_i$ and $x_o$ represent the original value and the scaled value of a pixel, respectively. The estimated fluorescent image stack was subsequently rescaled to its original range using $x_i = \frac{x_o}{255.0}(x_{99.99\%} - x_{0.01\%}) + x_{0.01\%}$. For those image stacks without ground truth values, the $\hat{x}_{0.01\%}$ and $\hat{x}_{0.99\%}$ can be estimated as the average of $x_{0.01\%}$ and $x_{99.99\%}$ related to the ground truth values in the training set.

Considering the limited number of image stacks in the three datasets described above, a 3-fold cross-validation (CV) approach was employed to train and validate the E-U-Nets after a few testing image stacks were held out for the E-U-Net testing. For a given dataset in which the testing stacks have been held out, the 3-fold CV approach involves randomly dividing all the stacks in the dataset into 3 folds of approximately equal size. The first two folds and the remaining one-fold were treated as a training set and a validation set to train and validate E-U-Nets, respectively. The procedure was repeated three times; each time, a different fold was treated as the validation set. The three procedures resulted in the validation of the E-U-Nets on each image stack. The trained E-U-Nets were finally tested on the held-out unseen testing samples. Details related to the cross-validation of E-U-Nets on the neuron, spheroids cell, and bead datasets are described below.

For the neuron dataset, two separate E-U-Nets were trained: one to translate phase images into each of the two-channel fluorescent images. The EfficientNet-B7 network was employed in the two E-U-Nets. The network architecture of the EfficientNet-B7 is shown in Fig. S6. Two neuron image stacks were held out as unseen testing data; the remaining 20 stacks were employed in the 3-fold CV process described above. In the 3-fold CV process, the 20 image stacks were randomly divided into three folds that contained 6, 7, and 7 image stacks, respectively. For each data split, the E-U-Nets were trained by minimizing a mean square error (MSE) loss function that measures the difference between the predicted fluorescent images and their corresponding ground truth values. The loss function was minimized by the use of an ADAM optimizer[47] with a learning rate of $5 \times 10^{-4}$, which was empirically determined. In each training iteration, a batch of paired 3 neighboring phase images and the corresponding central fluorescent image were sampled from the training image stacks, and then randomly cropped into patches of $512 \times 512$ pixels as training samples to train the networks. The batch size was set to 4. A decaying strategy was applied to the learning rate to mitigate the overfitting by multiplying the learning rate by 0.8 when the validation MSE loss did not decrease for consecutive 10 epochs. An epoch is a sequence of iterations that walk through all the image slices in the training set. The validation MSE loss was computed between the predicted fluorescent images and their ground truth values for validation images. In the network training, an early-stopping strategy was employed to determine the end of the network training. Specifically, at the end of each epoch, the being-trained E-U-Net model was evaluated by computing the average of Pearson correlation coefficients (PCCs) between the predicted fluorescent images and the related ground truth values. The network training stopped if the average validation PCC did not increase for 20 epochs as shown in Fig. S7. The two figures show the average training and validation stopping rule metric for training the two E-U-Nets respectively in one of the three training procedures of the 3-fold CV process. After the E-U-Nets were trained, the performances of the trained networks were evaluated on the validation set by computing the peak signal-to-noise ratio (PSNR) and PCC between the predicted fluorescent stacks and the related ground truth values. The 3-fold CV process resulted in validation results for each of the 20 stacks. These validation results were combined and are reported in SI Note 4. In addition, the E-U-Nets trained in the CV process were tested in the two unseen stacks. The corresponding Pearson correlation coefficients (PCCs) and PSNR are presented in SI Note 4.

For the spheroids cell dataset, two separate E-U-Nets were trained for each fluorescent channel. The EfficientNet-B7 network was employed as the encoder in the two trained E-U-Nets. Two spheroid cell image stacks were held out for E-U-Net testing; the remaining 19 stacks were randomly split into three folds that contain 6, 6, and 7 stacks, respectively, in the 3-fold CV process. The other training settings were the same as those described above for network training on the neuron dataset. The training and validation PCCs over epochs correspond to training the two E-U-Nets in one of the three training procedures of the 3-fold CV process are displayed in Fig. S8. The 3-fold CV results related to PSNR and PCC are reported in SI Note 4. The results tested on two unseen testing stacks are shown SI Note 4.

For the bead dataset, a single E-U-Net was built for the phase-to-fluorescent image translation. The EfficientNet-B0 was employed as the encoder in the E-U-Net. The architecture of the EfficientNet-B0 network is shown in Fig. S6. One of the bead image stacks was held out as an unseen testing stack for the E-U-Net testing; the remaining 17 bead stacks were randomly divided into three folds that each contains 5, 6, and 6 image stacks, respectively, for the 3-fold CV process. Paired images of size $128 \times 128$ pixels were employed for the E-U-Net training. The batch size was 32. The other training settings were the same as those for the network training on neuron and spheroid cell datasets, as described above. The training and validation stopping rule metric over epochs for one of the three training procedures of the 3-fold CV process are displayed in Fig. S9. The 3-fold CV results related to PSNR and PCC performances are reported in SI Note 4. The results on the unseen bead stack are shown in SI Note 4.

The E-U-Nets were implemented by use of the Python programming language with libraries including Python 3.6 and TensorFlow 1.14. The model training, validation, and testing were performed on an NVIDIA Tesla V100-GPU with 32 GB VRAM. The E-U-Net training on the neuron dataset and spheroid dataset took approximately 24 hours. The E-U-Net training on the bead dataset took approximately 2 hours. The inference time for a fluorescent image slice of 1744×1744 pixels was about approximately 400 $ms$.

**Neuron analysis:**

The volume of neurons was calculated from the ACM images using binary masks with background thresholding. The 3D dry mass distribution was generated with the multiplication of binary masks and the 3D dry mass distribution from the QPI images[48]. The 3D dry mass density is linearly related to the depth-resolved phase maps as

$$M(x, y, z) = \frac{\lambda}{2\pi\gamma\delta z} \phi(x, y, z), \qquad [2]$$

where $\lambda$ is the wavelength of the illumination and $\gamma \simeq 0.2$, the refractive increment, which lies within the range of 0.18–0.21 ml/g for most biological samples[49]. $\delta z$ is the z sampling, which for our LS-GLIM is ~ 1 $\mu m$. $\phi(x, y, z)$ is the measured phase image at each z-plane.

**Spheroid analysis:**

Three-dimensional semantic segmentation maps were generated from the estimated fluorescent signals corresponding to the RNASelect and 7-ADD stains by applying fixed thresholds for the entire data. This map of RNA and DNA-stained regions was further refined by assigning a "nucleoli" label to the RNA inside the DNA regions. To generate a map labeling the "spheroid", a threshold was applied to the quantitative phase signal after Hilbert demodulation[48]. Fields of view were acquired to contain a single spheroid, and phase values coincident with the assigned label ("nucleus", " spheroid") were totaled on a per-spheroid basis to report on the dry-mass and volume.

**Automated 3D cell counting:**

To segment our images into individual nuclei, we used a 3D variation of the marker-controlled watershed on the estimated DNA images[50]. We note that the ACM data lacked the unwanted pixel-level noise typically associated with photon-starved fluorescent images. This technique requires the image to be annotated into sample and background regions with a non-overlapping marker used to identify the cell. We performed 2D Hough voting which is used to identify the center of the nucleus in each z-slice, producing what resembles a curve through the z-dimension. To regularize our approach, we applied a 3x3 blur to correct for minor disconnects in our segmentation algorithm. The result of our watershed approach is a 3D volume with a unique label annotating each nucleus (Fig. 5a). To validate our method, we compared our results to a manual cell count performed in AMIRA. We obtained 142 cells counted automatically vs 136 cells counted manually (4 % error). The principal disagreement was due to undercounting touching cells. This procedure was implemented in MATLAB using the "imfindcirlces" and "watershed" commands.

**Sample preparation:**

**Hippocampal neuron preparation:**

All procedures involving animals were reviewed and approved by the Institutional Animal Care and Use Committee at the University of Illinois Urbana-Champaign and conducted per the guidelines of the U.S National Institute of Health (NIH). For our neuron imaging experiments, we used primary hippocampal neurons harvested from dissected hippocampi of Sprague-Dawley rat embryos at embryonic day 18. Dissociated hippocampal neurons were plated on multi-well plates (Cellvis, P06-20-1.5-N) that were pre-coated with poly-D-lysine (0.1 mg/ml; Sigma-Aldrich). Hippocampal neurons were incubated for three hours under the condition of 37°C and 5% $CO_2$ in a plating medium containing 86.55% MEM Eagle's with Earle's BSS (Lonza), 10% Fetal Bovine Serum (re-filtered, heat-inactivated; ThermoFisher), 0.45% of 20% (wt./vol.) glucose, 1x 100 mM sodium pyruvate (100x; Sigma-Aldrich), 1x 200 mM glutamine (100x; Sigma-Aldrich), and 1x Penicillin/ Streptomycin (100x; Sigma-Aldrich) in order to help attachment of neurons (300 cells/mm$^2$). The plating media was aspirated and replaced with maintenance media containing Neurobasal growth medium supplemented with B-27 (Invitrogen), 1% 200 mM glutamine (Invitrogen), and 1% penicillin/streptomycin (Invitrogen) and incubated for 10 days at 37 °C, in the presence of 5% $CO_2$. Hippocampal neurons were maintained for 14 days before performing immunostaining.

**Immunostaining protocol:**

Neurons were stained with antibodies for Tau (Abcam, ab80579) and MAP2 (Abcam ab32454) to localize axons and dendrites. Neurons were fixed with freshly prepared 4% paraformaldehyde (PFA) for 15 minutes following 0.5% Triton-X for 10 minutes and 2% bovine serum albumin (BSA, ThermoFisher) for 2 hours incubation in 4°C. Hippocampal neurons were incubated for 8 hours at 4°C with anti-Tau antibodies that were diluted to 1:250 in 5% BSA. After washing with

PBS, neurons were exposed for 8 hours at 4°C to goat anti-mouse secondary antibody (Abcam, ab205719) which was diluted to 1:500 in 5% BSA. Hippocampal neurons were then incubated in anti-MAP2 antibody (1:500 dilution) in 5% BSA for 8 hours, followed by goat antirabbit secondary antibody (Abcam, ab205718, 1:1000 dilution) in 5% BSA for 8 hours at 4°C.

**Liver cancer spheroid (HepG2 cells):**

Human hepatocarcinoma cells (HepG2, ATCC) were cultured in T-75 flasks with DMEM (Thermo) + 10% fetal bovine serum + 1% penicillin-streptomycin (Gibco) for 7 days and formed spontaneous pre-formed spheroids. The flasks were incubated at 37C and 5% $CO^2$. Media was replaced every 2-3 days. Spheroids were incubated with TrypLE Express (Thermo) for 10 minutes to detach pre-formed spheroids of approximately 100~200 μm in diameter from the culture flask. The passage number used was between 2-6.

Pre-formed spheroids were plated on poly-d-lysine coated glass-bottom dishes. The spheroids were incubated for 10 minutes to allow for attachment. Then, they were covered with a collagen hydrogel (bovine collagen type 1, Advanced Biomatrix). The cells were incubated for 3 days to allow for cellular reorganization into a regular spheroidal shape. The spheroids were first fixed in a 1:1 ratio of methanol: acetone at -4°C for 20 minutes. Cells fixed using this method do not need an additional permeabilization step due to the acetone. The cell nucleus was stained using 7-aminoactinomycin D (7-AAD red, 6163, ThermoFisher) by adding 1μL of the stock stain into 1mL of PBS. The cell RNA was stained using SYTO RNASelect Green (S32703, ThermoFisher) by first creating a 5μM working solution and then adding 100μL of the working solution to 900μL of PBS. The samples were stained at room temperature for 30 minutes before rinsing once. Two kinds of samples in PBS or hydrogel were imaged after staining.

**Data availability:**

The data that support the findings of this study are available from the corresponding author upon reasonable request.

**Code availability:**

The codes that support the findings of this study are available from the corresponding author upon reasonable request.

**Acknowledgments:**

This work is supported by the National Science Foundation (CBET0939511 STC, NRT-UtB 1735252), the National Institute of General Medical Sciences (GM129709), and the National Cancer Institute (CA238191).

**Author contributions:**

X.C., M.E.K., and G.P. conceived the project. X.C. and M.E.K. designed the experiments. X.C. and M.E.K. built the system. X.C. performed imaging. S.H. trained the machine learning network. X.C. and M.E.K. analyzed the data. G.T & H.J.C. provided neurons. Y.J.L. cultured neurons and performed immunocytochemistry. K.M.S. & H.K. provided spheroids. C.H., X.C., and G.P.

derived the theoretical model. X.C., M.E.K., S.H., C.H., and G.P. wrote the manuscript. M.A. supervised the AI work. G.P. supervised the project.

**Competing interests:**

G.P. has a financial interest in Phi Optics, Inc., a company developing quantitative-phase-imaging technology for materials and life science applications. The remaining authors declare that they have no conflict of interest.

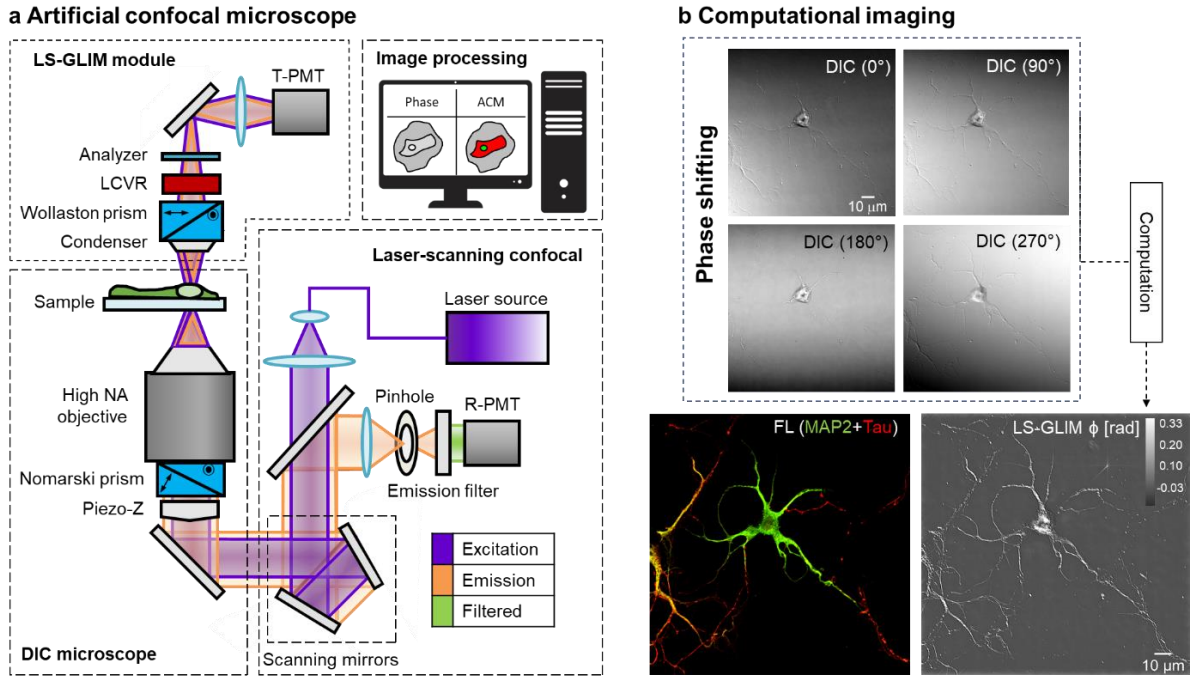

Figure 1. **a,** The ACM system consists of a laser-scanning confocal assembly, differential interference contrast (DIC) microscope, and LS-GLIM module. The quantitative phase imaging was conducted with the green laser line (488 nm) of the confocal excitation. The interferogram was recorded at each point in the scan by the transmission-PMT (T-PMT). The fluorescence images were captured by the reflection-PMT of the confocal module. **b,** Four phase-shifting frames are recorded and used to reconstruct the quantitative phase image. The confocal fluorescence image (FL) serves as the ground truth, and the phase image (LS-GLIM) is the input for the network training.

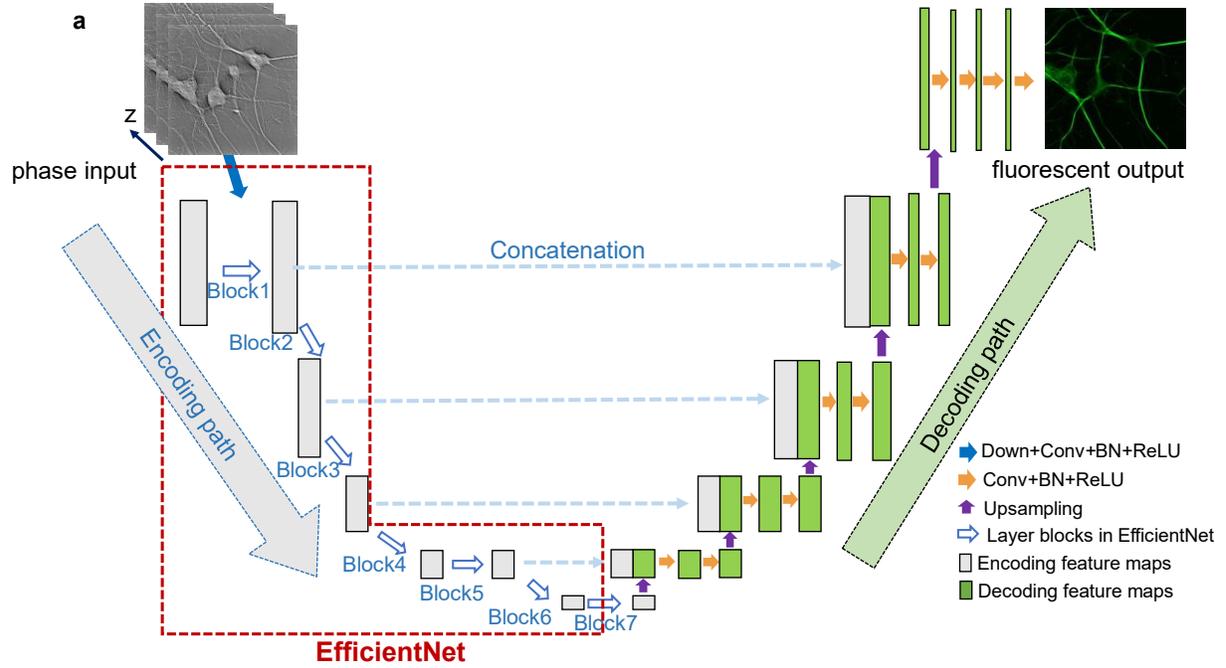

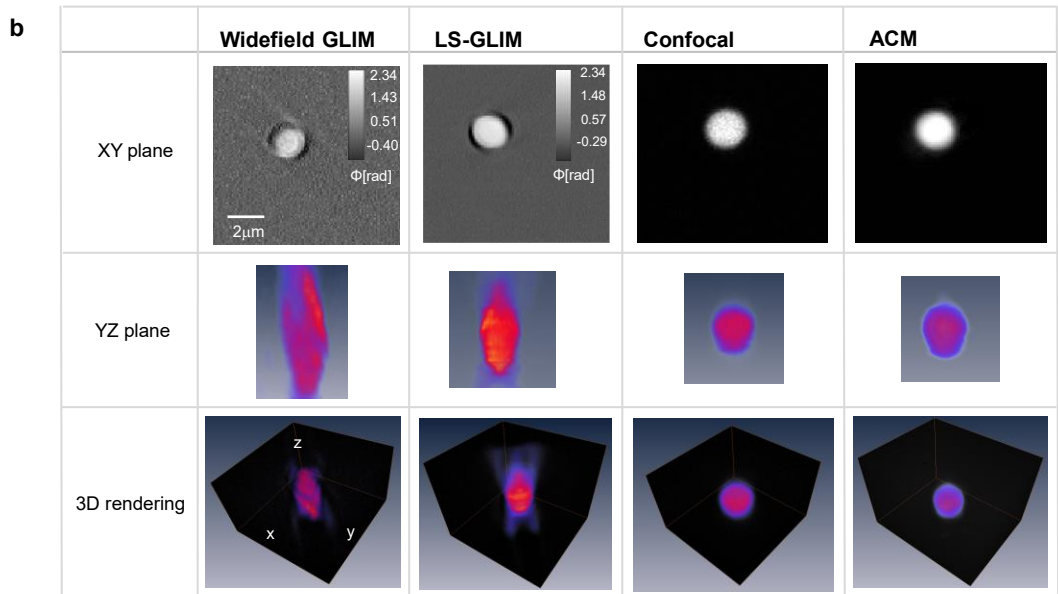

Figure 2. **a,** Network architecture for translating phase images into confocal fluorescence signals. It is a U-Net variant that uses an EfficientNet as the encoder. The input of the Efficient-Unet consists of 3 adjacent quantitative phase images along the z-axis, and its output is the corresponding middle fluorescent slice. **b,** Comparison of 2 μm bead tomograms in widefield, LS-GLIM, confocal, and ACM, as indicated. The elongation of the beads in widefield and LS-GLIM is due to the missing frequencies in the transmission geometry. On the other hand, the predicted ACM images replicate the confocal sectioning and resolution.

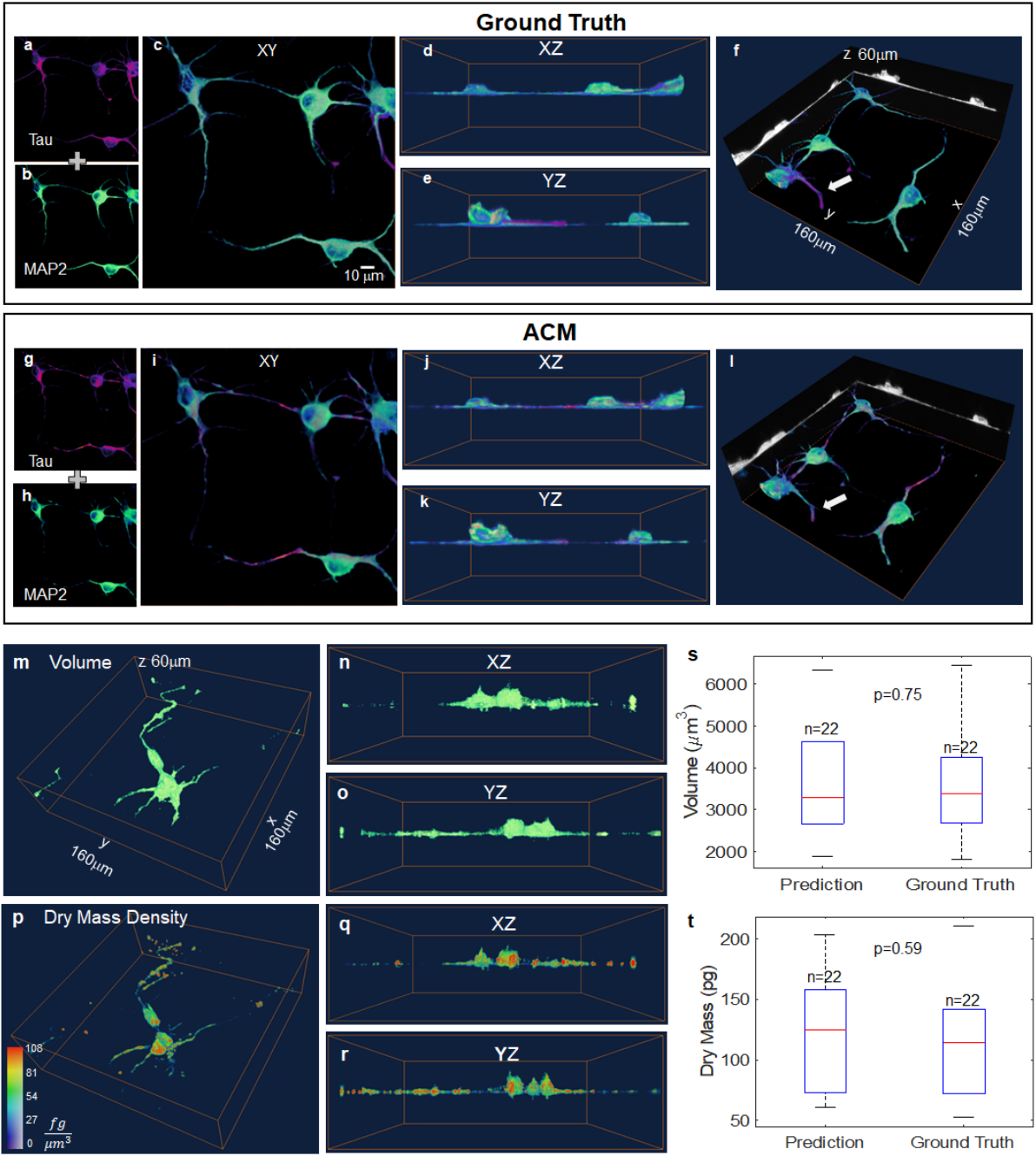

Figure 3. 2D comparison of ground truth from confocal fluorescence (**a,** Tau, **b,** MAP2) and predicted fluorescence (**g,** Tau, **h,** MAP2). 3D comparison of ground truth from confocal fluorescence (**c,** XY, **d,** XZ, **e,** YZ, **f,** 3D tomogram) and predicted fluorescence (**i,** XY, **j,** XZ, **k,** YZ, **l,** 3D tomogram). 3D MAP2 associated volume (**m,** 3D volume, **n,** YZ **o,** XZ) and dry mass distribution (**p,** 3D dry mass density **q,** YZ **r,** XZ). The box plots of volume (**s**) and dry mass calculation (**t**) based on ground truth (GT) and predictions (PR) for 10 days *in vitro* (DIV) neurons.

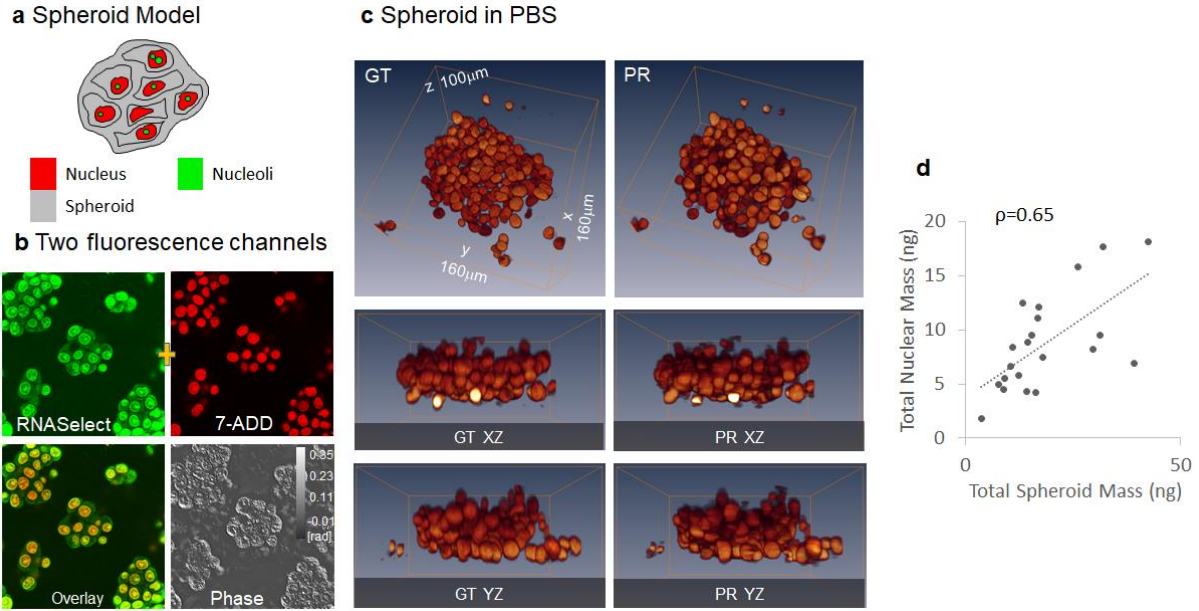

Figure 4. Label-free intracellular segmentation in turbid spheroids. **a & b,** Cellular compartments were stained using RNA and DNA sensitives stains:: DNA is used to identify the nucleus and dense concentrations of RNA inside the nucleus are associated nucleoli. **c,** A deep convolutional neural network was trained to estimate the two fluorescent stains from the label-free images (rendering of recorded and estimated DNA stain, 40x/1.3). For all 20 spheroids we calculated the nuclear dry-mass and volumes generated from the imputed signal. **d,** Total nuclear dry-mass tracks closely with total spheroid mass, Person correlation coefficient ρ=+0.65 (the slope of the linear fit is 0.42).

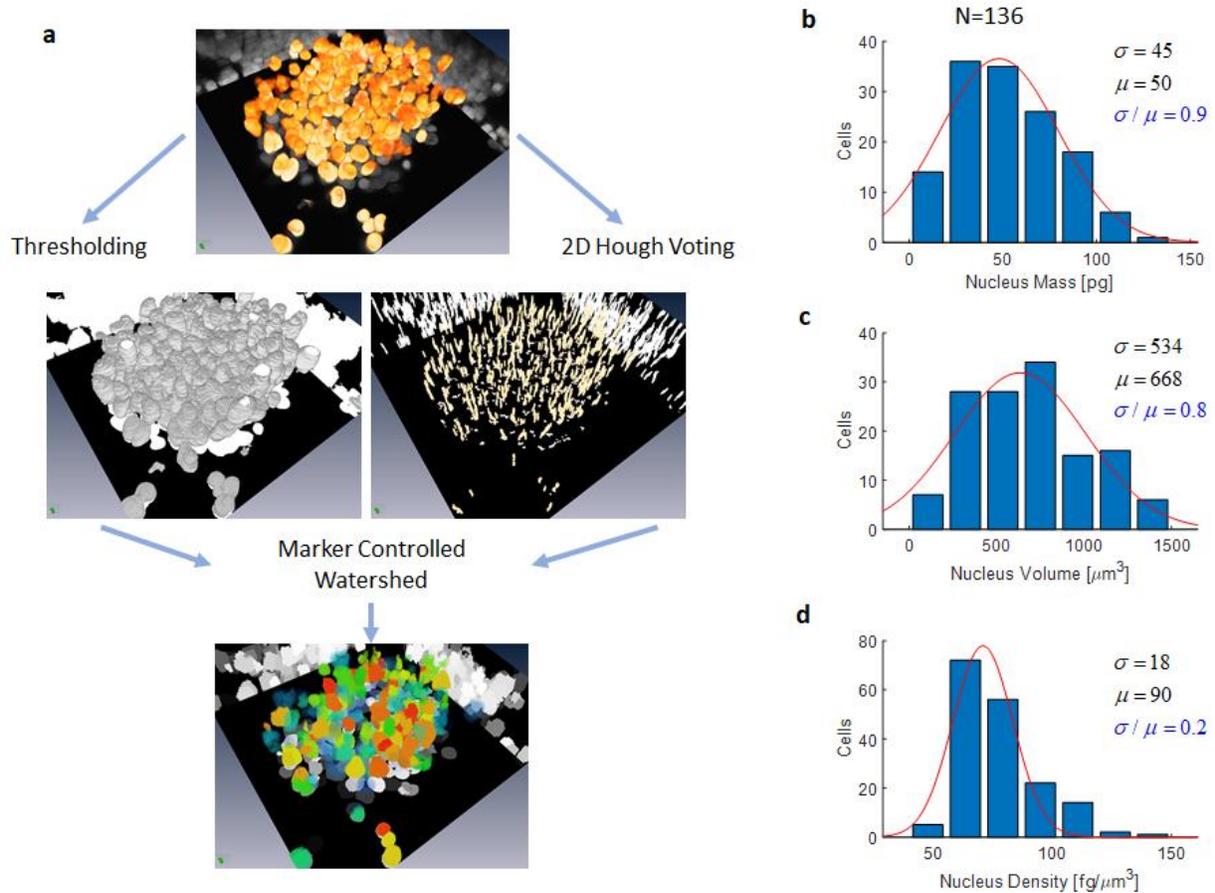

Figure 5. Automated segmentation of cells inside spheroids. **a,** Instance segmentation of spheroids was performed by 3D marker-controlled watershed on the estimated DNA signal, with markers being determined through 2D Hough voting on a per-z-slice basis. The result of the Hough voting is a volume with a unique marker on the spheroid, which resembles a column tracking the center of the nucleus through the focus. The result of watershed is a 3D volume with a unique label for each nucleus within the spheroid, which enables the calculation of parameters on individual cells. **b,** Distribution of nuclear dry mass. **c,** Distribution of nuclear volume. **d,** Distribution of dry mass density. The standard deviation ($\sigma$), mean ($\mu$), and their ratios are indicated for each plot. Note that the nuclear mass density (**d**) has a significantly narrower distribution compared to nuclear mass (**b**), and volume (**c**), as indicated by the $\sigma/\mu$ ratio.

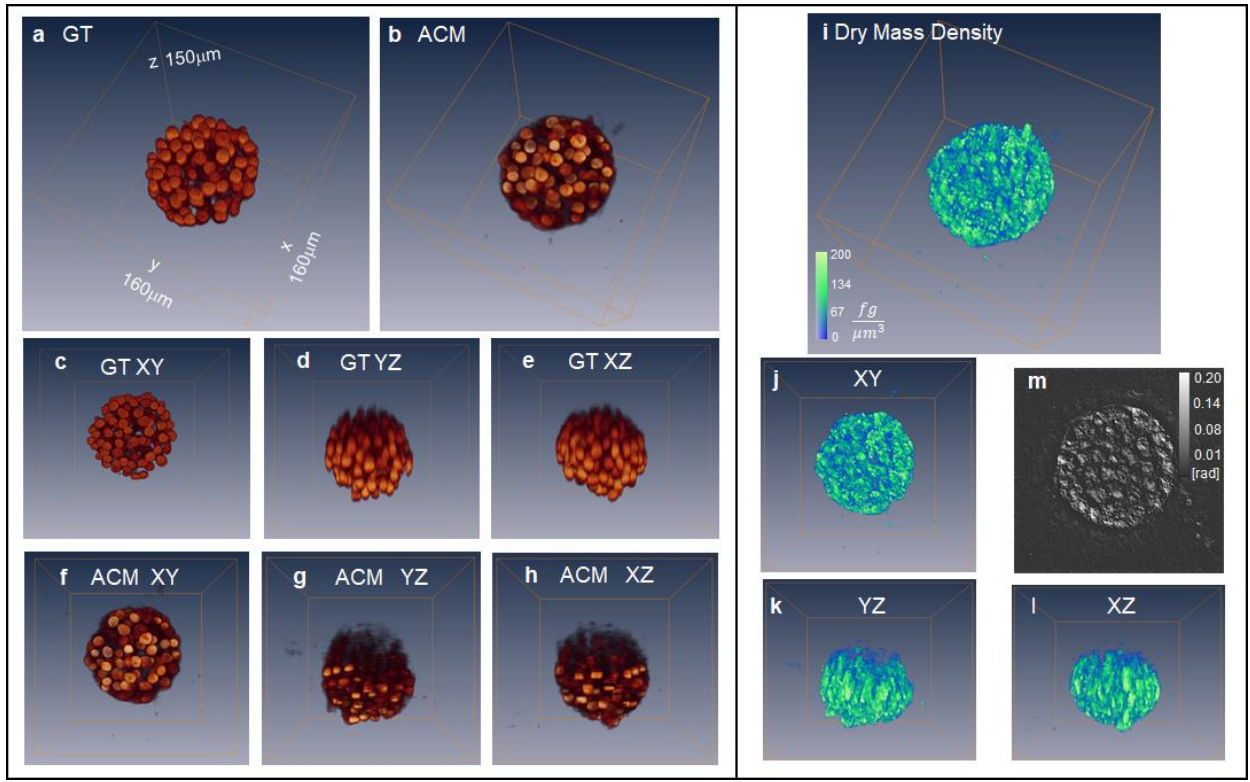

Figure 6. A confocal image of DNA-tagged spheroid in hydrogel as ground truth (GT) data for the network. Comparison of ground truth (**a,** 3D tomogram, **c,** XY, **d,** YZ, **e,** XZ) and prediction (**b,** 3D tomogram, **f,** XY, **g,** YZ, **h,** XZ) of nuclei of a liver cancer spheroid (HepG2) in hydrogel. **i,** 3D dry mass density of the spheroid's nuclei (**j,** XY, **k,** YZ, **m,** XZ). **l,** Quantitative phase image for the spheroid in hydrogel.

Supplemental information for

**Artificial confocal microscopy for deep label-free imaging**

*Xi Chen[1], Mikhail E. Kandel[1,2], Shenghua He[3], Chenfei Hu[1,4], Young Jae Lee[1,5], Kathryn Sullivan[6], Gregory Tracy[7], Hee Jung Chung[5,7], Hyun Joon Kong[1,6,8,9], Mark Anastasio[1,6], Gabriel Popescu[1,4,6,9]*

# Supplementary Note 1: LCVR calibration procedure

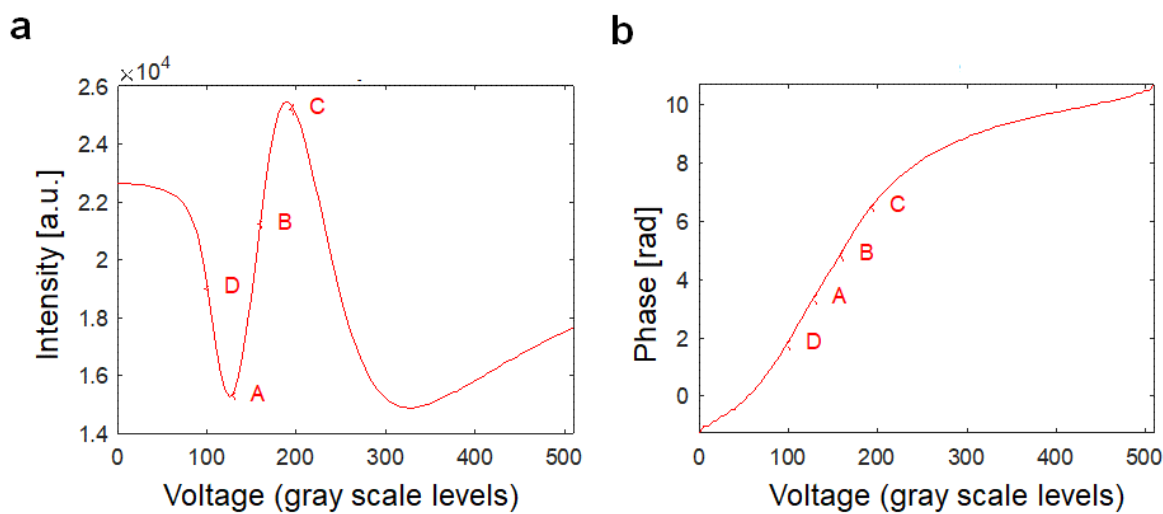

*Fig. S1.* LCVR calibration procedure. The relationship between the LCVR voltage and phase shift is determined by acquiring intensity images with increasing modulator voltage (a). The phase shift imparted by the modulator is obtained through a Hilbert transform (b). A, B, C, D indicate frames at 90° phase shifts used for the four-frame shifting interferometry.

# Supplementary Note 2: Phase sensitivity comparison between widefield and laser-scanning gradient light interference microscopy

We compare the phase sensitivity of widefield gradient light interference microscopy and laser-scanning interference microscopy by measuring the same background region on both channels and computing the standard deviation of the noise. A 485nm/20nm filter was used in the illumination path of widefield gradient light interference microscopy to match the laser line (488 nm) of laser-scanning interference microscopy. The histograms of the intensity of the four frames are matched on both channels. We found that the standard deviation of the background phase from LS-GLIM is one-fifth of that from (widefield) GLIM, demonstrating the superior sensitivity of the laser-scanning method. The significantly lower noise signature associated with LS-GLIM is due to the more sensitive detectors and absence of spatial crosstalk.

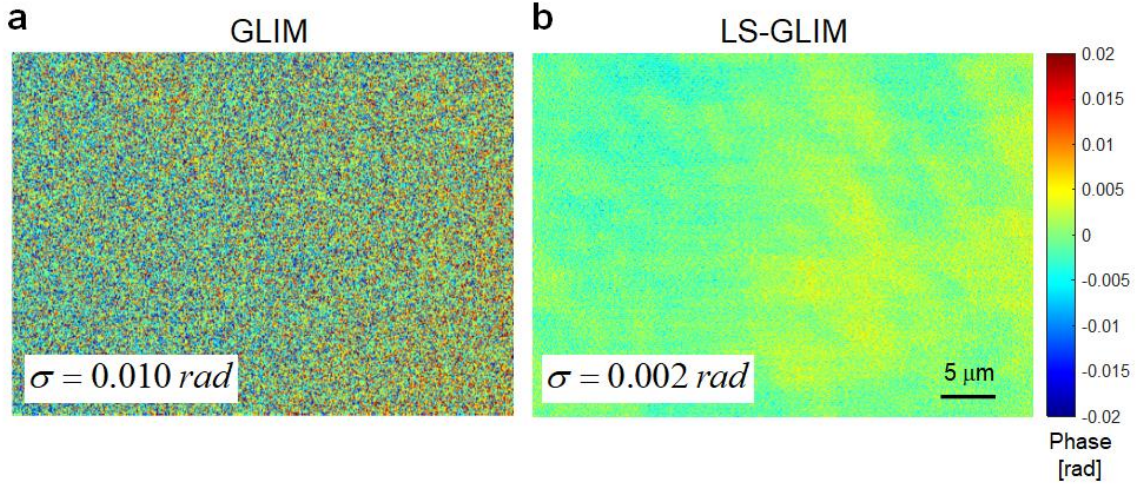

*Fig. S2. Phase sensitivity comparison between (widefield) GLIM (a) and LS-GLIM (b). The spatially-averaged standard deviation of the phase noise, σ, is indicated for each case.*

**Supplementary Note 3: 3D image formation in LS-GLIM**

Here we derive a physical model for 3D image formation in LS-GLIM. This model takes into account the scanning, focused illumination, light scattering from the sample, and the interferometric detection, to derive an expression for the 3D transfer function of the instrument.

The propagation of the total field $U(\mathbf{r})$ through a static scattering medium is governed by the inhomogeneous Helmholtz equation [51],

$$\nabla^2 U(\mathbf{r}) + n_0^2 \beta_0^2 U(\mathbf{r}) = -n_0^2 \beta_0^2 \chi(\mathbf{r}) U(\mathbf{r}), \quad (1)$$

where $\beta_0 = \dfrac{\omega}{c}$ is the wavenumber in vacuum, $n_0$ is the refractive index of the surrounding medium, $\omega$ is the frequency of the light, $c$ is the speed of light in vacuum, and $\chi = n^2(\mathbf{r}) - n_0^2$ is the scattering potential of the sample. We can split the total field into incident ($U_i$), and scattered field ($U_s$), which satisfy the following Helmholtz equations, respectively, as

$$\nabla^2 U_i(\mathbf{r}) + \beta^2 U_i(\mathbf{r}) = 0, \quad (2)$$

$$\nabla^2 U_s(\mathbf{r}) + \beta^2 U_s(\mathbf{r}) = -\beta_0^2 \chi(\mathbf{r}) U_i(\mathbf{r}), \quad (3)$$

where $\beta = n_0 \beta_0$ is the wavenumber in the surrounding medium.

LS-GLIM shares the same laser source as the confocal fluorescence microscope. The incident field propagates through the objective to reach the sample plane. Let us consider that the circular aperture $A_i$ of the illumination field is of the form (Fig. S3)

$$A_i(\mathbf{k}_\perp) = \begin{cases} 1 & |\mathbf{k}_\perp| \leq \beta_0 \mathrm{NA}_o \\ 0 & |\mathbf{k}_\perp| > \beta_0 \mathrm{NA}_o \end{cases}, \tag{4}$$

where $\mathrm{NA}_o$ is the numerical aperture (NA) of the objective.

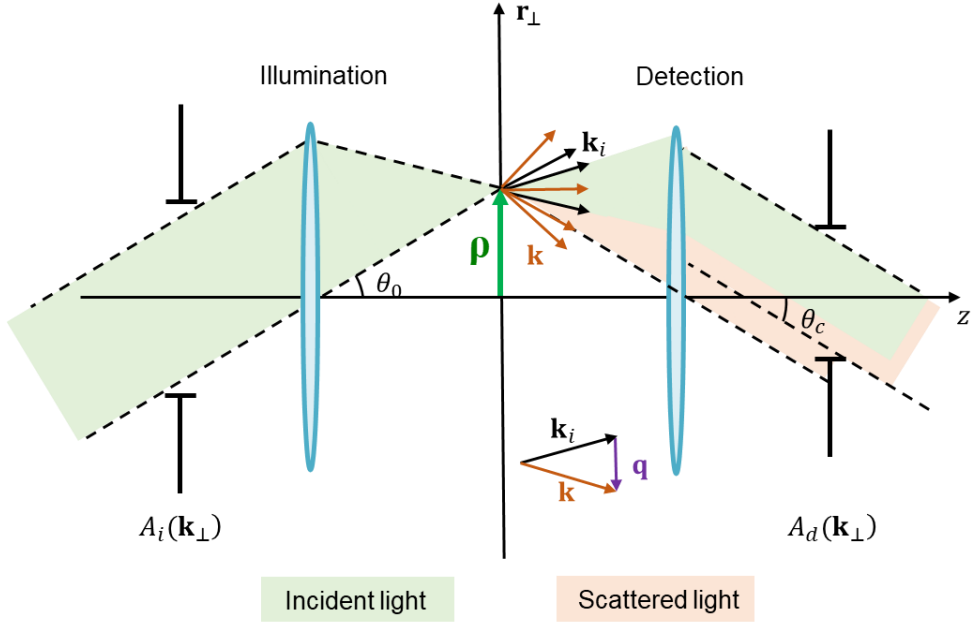

*Fig. S3. Illustration of the laser-scanning interference microscopy light-path and scattering event.*

The incident field is focused at the sample plane into a point at coordinate $(\mathbf{r}_\perp, z)$, and it consists of plane waves originating in the Fourier domain $\mathbf{k} = (\mathbf{k}_\perp, k_z)$. Let us first consider one plane wave in the sample plane [52],

$$U_{i0}(\mathbf{r}_\perp, z) = A_i(\mathbf{k}'_{i\perp}) e^{i\mathbf{k}'_{i\perp} \cdot \mathbf{r}_\perp} e^{i\gamma(\mathbf{k}'_{i\perp})z}, \tag{5}$$

where $\gamma(\mathbf{k}'_{i\perp}) = \sqrt{\beta^2 - k'^2_{i\perp}}$, and $k'_{i\perp} = |\mathbf{k}'_{i\perp}|$. Taking the Fourier transform with respect to $\mathbf{r}_\perp$, we have

$$U_{i0}(\mathbf{k}_\perp, z) = A_i(\mathbf{k}_{i\perp}) \delta(\mathbf{k}_\perp - \mathbf{k}'_{i\perp}) e^{i\gamma(\mathbf{k}'_{i\perp})z}. \tag{6}$$

To calculate the scattered field, we take the 3D Fourier transform of Eq. (3). Thus, the forward scattered field ($z > 0$) under the first-order Born approximation is [53]

$$U_s(\mathbf{k}) = \frac{-\beta_0^2}{2\gamma(\mathbf{k}_\perp)} \frac{1}{\gamma(\mathbf{k}_\perp) - k_z} \left[ \chi(\mathbf{k}) \circledv_\mathbf{k} U_{i0}(\mathbf{k}) \right], \tag{7}$$

where $\circledv_\mathbf{k}$ denotes the 3D convolution in the $\mathbf{k}$-domain. We can express Eq. (7) in the $(\mathbf{k}_\perp, z)$ domain by taking the inverse Fourier transform along $k_z$, and by using the fact that $e^{i\gamma(\mathbf{k}_\perp)z} \circledv_z f(z) = e^{i\gamma(\mathbf{k}_\perp)z} f(\gamma)$ (see Chapter 4 in Ref. [54]), we obtain

$$U_s(\mathbf{k}_\perp, z) = \frac{i\beta_0^2}{2\gamma(\mathbf{k}_\perp)} e^{i\gamma(\mathbf{k}_\perp)z} \left[ \chi(\mathbf{k}) \circledv U_{i0}(\mathbf{k}) \right]\bigg|_{k_z = \gamma(\mathbf{k}_\perp)}. \tag{8}$$

Considering the situation where $U_s$ is produced by the single plane wave in Eq. (6), we can evaluate the scattered field as

$$U_s(\mathbf{k}_\perp, z) = \frac{i\beta_0^2}{2\gamma(\mathbf{k}_\perp)} A_i(\mathbf{k}_{i\perp}'') e^{i\gamma(\mathbf{k}_\perp)z} \chi\left[ \mathbf{k}_\perp - \mathbf{k}_{i\perp}'', \gamma(\mathbf{k}_\perp) - \gamma(\mathbf{k}_{i\perp}'') \right]. \tag{9}$$

where $\mathbf{k}_{i\perp}''$ denotes the wavevector of incident light propagating in different directions. At the detector plane, both $U_s$ and $U_i$ are filtered by the detection aperture, $A_d(\mathbf{k}_\perp)$. In LS-GLIM, the condenser aperture plays the role of the detection pupil, i.e., $A_d$ for a circular aperture is in the form

$$A_d(\mathbf{k}_\perp) = \begin{cases} 1 & |\mathbf{k}_\perp| \leq \beta_0 \mathrm{NA}_c \\ 0 & |\mathbf{k}_\perp| > \beta_0 \mathrm{NA}_c \end{cases}, \tag{10}$$

$\mathrm{NA}_c$ is the numerical aperture of the condenser for laser-scanning interference microscopy. The fields at the detector plane are

$$U_s(\mathbf{k}_\perp, z) = \frac{i\beta_0^2}{2\gamma(\mathbf{k}_\perp)} e^{i\gamma(\mathbf{k}_\perp)z} A_i(\mathbf{k}_{i\perp}'') A_d(\mathbf{k}_\perp) \chi\left[ \mathbf{k}_\perp - \mathbf{k}_{i\perp}'', \gamma(\mathbf{k}_\perp) - \gamma(\mathbf{k}_{i\perp}'') \right], \tag{11}$$

$$U_i(\mathbf{k}_\perp, z) = A_i(\mathbf{k}_{i\perp}') A_d(\mathbf{k}_\perp) \delta(\mathbf{k}_\perp - \mathbf{k}_{i\perp}') e^{i\gamma(\mathbf{k}_{i\perp}')z}. \tag{12}$$

Next, we calculate the cross-spectral density between $U_s$ and $U_i$, and then integrate the results over all possible illumination angles. The cross-spectral density in $(\mathbf{k}_\perp, z)$ domain is

$$W(\mathbf{k}_\perp, z) = U_s(\mathbf{k}_\perp, z) \circledv_{\mathbf{k}_\perp} U_i^*(-\mathbf{k}_\perp, z)$$

$$= \left[ i\frac{\beta_0^2}{2} A_i(\mathbf{k}_{i\perp}'') A_i(\mathbf{k}_{i\perp}') \right] \left\{ A_d(\mathbf{k}_\perp) \frac{e^{i[\gamma(\mathbf{k}_\perp) - \gamma(\mathbf{k}_{i\perp}')]z}}{\gamma(\mathbf{k}_\perp)} \chi\left[\mathbf{k}_\perp - \mathbf{k}_{i\perp}'', \gamma(\mathbf{k}_\perp) - \gamma(\mathbf{k}_{i\perp}'')\right] \right\} \quad (13)$$

$$\circledv_{\mathbf{k}_\perp} \left\{ A_d^*(-\mathbf{k}_\perp) \left[ \delta(\mathbf{k}_\perp + \mathbf{k}_{i\perp}') \right] \right\}.$$

In general, $A_d(\mathbf{k}_\perp)$ is Hermitian, so we can replace $A_d^*(-\mathbf{k}_\perp)$ in Eq. (13) with $A_d(\mathbf{k}_\perp)$. In addition, invoking the properties of $\delta$ function,

$$A_d(\mathbf{k}_\perp)\left[\delta(\mathbf{k}_\perp + \mathbf{k}_{i\perp}')\right] = A_d(-\mathbf{k}_{i\perp}')\left[\delta(\mathbf{k}_\perp + \mathbf{k}_{i\perp}')\right], \quad (14)$$

$$f(\mathbf{k}_\perp) \circledv_{\mathbf{k}_\perp} \delta(\mathbf{k}_\perp + \mathbf{k}_{i\perp}') = f(\mathbf{k}_\perp + \mathbf{k}_{i\perp}'). \quad (15)$$

Equation (13) becomes,

$$W(\mathbf{k}_\perp, z) = \left[ i\frac{\beta_0^2}{2} A_i(\mathbf{k}_{i\perp}'') A_i(\mathbf{k}_{i\perp}') A_d(\mathbf{k}_{i\perp}') \right] \cdot$$

$$\left\{ A_d(\mathbf{k}_\perp + \mathbf{k}_{i\perp}') \frac{e^{i[\gamma(\mathbf{k}_\perp + \mathbf{k}_{i\perp}') - \gamma(\mathbf{k}_{i\perp}'')]z}}{\gamma(\mathbf{k}_\perp + \mathbf{k}_{i\perp}')} \chi\left[\mathbf{k}_\perp - \mathbf{k}_{i\perp}'' + \mathbf{k}_{i\perp}', \gamma(\mathbf{k}_\perp + \mathbf{k}_{i\perp}') - \gamma(\mathbf{k}_{i\perp}'')\right] \right\}. \quad (16)$$

The variable in the denominator, $\gamma(\mathbf{k}_\perp + \mathbf{k}_{i\perp}')$, can be approximated as $n_0\beta_0$ with $n_0$ the refractive of the background, and the result of Eq. (16) can be simplified

$$W(\mathbf{k}_\perp, z) = i\frac{\beta_0}{2n_0}\left[A_i(\mathbf{k}_{i\perp}'')A_i(\mathbf{k}_{i\perp}')A_d(\mathbf{k}_{i\perp}')\right] \cdot$$

$$\left\{ A_d(\mathbf{k}_\perp + \mathbf{k}_{i\perp}') e^{i[\gamma(\mathbf{k}_\perp + \mathbf{k}_{i\perp}') - \gamma(\mathbf{k}_{i\perp}'')]z} \chi\left[\mathbf{k}_\perp - \mathbf{k}_{i\perp}'' + \mathbf{k}_{i\perp}', \gamma(\mathbf{k}_\perp + \mathbf{k}_{i\perp}') - \gamma(\mathbf{k}_{i\perp}'')\right] \right\}. \quad (17)$$

Taking a Fourier transform with respect to $z$, the cross-spectral density in the $\mathbf{k}$ domain is obtained

$$W(\mathbf{k}) = i\frac{\beta_0}{2n_0}\left[A_i(\mathbf{k}_{i\perp}'')A_i(\mathbf{k}_{i\perp}')A_d(\mathbf{k}_{i\perp}')\right] \cdot$$

$$\left\{ A_d(\mathbf{k}_\perp + \mathbf{k}_{i\perp}') \chi\left[\mathbf{k}_\perp - \mathbf{k}_{i\perp}'' + \mathbf{k}_{i\perp}', \gamma(\mathbf{k}_\perp + \mathbf{k}_{i\perp}') - \gamma(\mathbf{k}_{i\perp}'')\right] \delta\left[k_z - \gamma(\mathbf{k}_\perp + \mathbf{k}_{i\perp}') + \gamma(\mathbf{k}_{i\perp}'')\right] \right\}. \quad (18)$$

An arbitrary scattering potential, $\chi$ can be written as a convolution in the spatial domain, namely,

$$\chi(\mathbf{r}) = \int \chi(\boldsymbol{\varepsilon})\delta(\mathbf{r} - \boldsymbol{\varepsilon})d\boldsymbol{\varepsilon}. \quad (19)$$

Thus, $\chi\left[\mathbf{k}_\perp - \mathbf{k}_{i\perp}^{''} + \mathbf{k}_{i\perp}^{'}, \gamma\left(\mathbf{k}_\perp + \mathbf{k}_{i\perp}^{'}\right) - \gamma\left(\mathbf{k}_{i\perp}^{''}\right)\right]$ can be expressed as

$$\chi\left[\mathbf{k}_\perp - \mathbf{k}_{i\perp}^{''} + \mathbf{k}_{i\perp}^{'}, \gamma\left(\mathbf{k}_\perp + \mathbf{k}_{i\perp}^{'}\right) - \gamma\left(\mathbf{k}_{i\perp}^{''}\right)\right] = \int \chi(\varepsilon) d\varepsilon \int \delta(\mathbf{r} - \varepsilon) e^{i\left\{\left(\mathbf{k}_\perp - \mathbf{k}_{i\perp}^{''} + \mathbf{k}_{i\perp}^{'}\right)\mathbf{r}_\perp + \left[\gamma\left(\mathbf{k}_\perp + \mathbf{k}_{i\perp}^{'}\right) - \gamma\left(\mathbf{k}_{i\perp}^{''}\right)\right]z\right\}} d\mathbf{r}$$

$$= \int \chi(\varepsilon) e^{i\left\{\left(\mathbf{k}_\perp - \mathbf{k}_{i\perp}^{''} + \mathbf{k}_{i\perp}^{'}\right)\varepsilon_\perp + \left[\gamma\left(\mathbf{k}_\perp + \mathbf{k}_{i\perp}^{'}\right) - \gamma\left(\mathbf{k}_{i\perp}^{''}\right)\right]\varepsilon_z\right\}} d\varepsilon,$$

(20)

where $\varepsilon_\perp$ and $\varepsilon_z$ are the $\varepsilon$ coordinates in the transverse and axial directions, respectively. Applying the result of Eq. (20), Eq. (18) can thus be rewritten as

$$W(\mathbf{k}) = \int i \frac{\beta_0}{2n_0} \left[A_i\left(\mathbf{k}_{i\perp}^{''}\right) A_i\left(\mathbf{k}_{i\perp}^{'}\right) A_d\left(\mathbf{k}_{i\perp}^{'}\right)\right] \cdot$$

$$\left\{A_d\left(\mathbf{k}_\perp + \mathbf{k}_{i\perp}^{'}\right) \chi(\varepsilon) e^{i\left\{\left(\mathbf{k}_\perp - \mathbf{k}_{i\perp}^{''} + \mathbf{k}_{i\perp}^{'}\right)\varepsilon_\perp + \left[\gamma\left(\mathbf{k}_\perp + \mathbf{k}_{i\perp}^{'}\right) - \gamma\left(\mathbf{k}_{i\perp}^{''}\right)\right]\varepsilon_z\right\}} \delta\left[k_z - \gamma\left(\mathbf{k}_\perp + \mathbf{k}_{i\perp}^{'}\right) + \gamma\left(\mathbf{k}_{i\perp}^{''}\right)\right]\right\} d\varepsilon.$$

(21)

The photodetector collects light produced by all illumination and scattering angles. Thus, the result of Eq. (21) needs to integrate over $\mathbf{k}_{i\perp}^{'}$, $\mathbf{k}_{i\perp}^{''}$, $\mathbf{k}_\perp$, and $k_z$, meaning, the detected signal is

$$s = \iiint W(\mathbf{k}) d\mathbf{k}_{i\perp}^{''} d\mathbf{k}_{i\perp}^{'} d\mathbf{k}_\perp dk_z$$

$$= i \frac{\beta_0}{2n_0} \int d\mathbf{k}_{i\perp}^{''} \int d\mathbf{k}_{i\perp}^{'} \int d\mathbf{k}_\perp \int dk_z \int d\varepsilon \left[A_i\left(\mathbf{k}_{i\perp}^{''}\right) A_i\left(\mathbf{k}_{i\perp}^{'}\right) A_d\left(\mathbf{k}_{i\perp}^{'}\right)\right] \cdot \quad (22)$$

$$\left\{A_d\left(\mathbf{k}_\perp + \mathbf{k}_{i\perp}^{'}\right) \chi(\varepsilon) e^{i\left[\left(\mathbf{k}_\perp - \mathbf{k}_{i\perp}^{''} + \mathbf{k}_{i\perp}^{'}\right)\varepsilon_\perp + \left[\gamma\left(\mathbf{k}_\perp + \mathbf{k}_{i\perp}^{'}\right) - \gamma\left(\mathbf{k}_{i\perp}^{''}\right)\right]\varepsilon_z\right]} \delta\left[k_z - \gamma\left(\mathbf{k}_\perp + \mathbf{k}_{i\perp}^{'}\right) + \gamma\left(\mathbf{k}_{i\perp}^{''}\right)\right]\right\}.$$

Let us compute the result of Eq. (22) step-by-step. First, let us adjust the sequence and start with the integration over $\mathbf{k}_{i\perp}^{''}$,

$$s = i \frac{\beta_0}{2n_0} \int d\varepsilon \int d\mathbf{k}_{i\perp}^{'} \int d\mathbf{k}_\perp \int dk_z \chi(\varepsilon)$$

$$\left[A_i\left(\mathbf{k}_{i\perp}^{'}\right) A_d\left(\mathbf{k}_{i\perp}^{'}\right)\right] A_d\left(\mathbf{k}_\perp + \mathbf{k}_{i\perp}^{'}\right) e^{i\left[\left(\mathbf{k}_\perp + \mathbf{k}_{i\perp}^{'}\right)\varepsilon_\perp + \gamma\left(\mathbf{k}_\perp + \mathbf{k}_{i\perp}^{'}\right)\varepsilon_z\right]} \delta\left[k_z - \gamma\left(\mathbf{k}_\perp + \mathbf{k}_{i\perp}^{'}\right) + \gamma\left(\mathbf{k}_{i\perp}^{''}\right)\right] \quad (23)$$

$$\int A_i\left(\mathbf{k}_{i\perp}^{''}\right) e^{-i\left[\mathbf{k}_{i\perp}^{''}\varepsilon_\perp + \gamma\left(\mathbf{k}_{i\perp}^{''}\right)\varepsilon_z\right]} d\mathbf{k}_{i\perp}^{''}.$$

It is not hard to find that

$$\int A_i\left(\mathbf{k}_{i\perp}^{''}\right) e^{-i\left[\mathbf{k}_{i\perp}^{''} \cdot \varepsilon_\perp + \gamma\left(\mathbf{k}_{i\perp}^{''}\right)\varepsilon_z\right]} d\mathbf{k}_{i\perp}^{''} = U_i^*(\varepsilon). \quad (24)$$

Equation (24) represents the conjugate incident field evaluated at $\varepsilon$. As a result, we can simplify the expression of Eq. (23) to be

$$s = i\frac{\beta_0}{2n_0} \int d\boldsymbol{\varepsilon} \int d\mathbf{k}'_{i\perp} \int d\mathbf{k}_\perp \int dk_z \chi(\boldsymbol{\varepsilon}) U_i^*(\boldsymbol{\varepsilon})$$
$$\left[ A_i(\mathbf{k}'_{i\perp}) A_d(\mathbf{k}'_{i\perp}) \right] A_d(\mathbf{k}_\perp + \mathbf{k}'_{i\perp}) e^{i\left[ (\mathbf{k}_\perp + \mathbf{k}'_{i\perp})\varepsilon_\perp + \gamma(\mathbf{k}_\perp + \mathbf{k}'_{i\perp})\varepsilon_z \right]} \delta\left[ k_z - \gamma(\mathbf{k}_\perp + \mathbf{k}'_{i\perp}) + \gamma(\mathbf{k}'_{i\perp}) \right]. \quad (25)$$

Next, we re-arrange the integration order one more time and compute the integration over $k_z$,

$$s = i\frac{\beta_0}{2n_0} \int d\boldsymbol{\varepsilon} \int d\mathbf{k}'_{i\perp} \int d\mathbf{k}_\perp \chi(\boldsymbol{\varepsilon}) U_i^*(\boldsymbol{\varepsilon})$$
$$\left[ A_i(\mathbf{k}'_{i\perp}) A_d(\mathbf{k}'_{i\perp}) \right] A_d(\mathbf{k}_\perp + \mathbf{k}'_{i\perp}) e^{i\left[ (\mathbf{k}_\perp + \mathbf{k}'_{i\perp})\varepsilon_\perp + \gamma(\mathbf{k}_\perp + \mathbf{k}'_{i\perp})\varepsilon_z \right]} \quad (26)$$
$$\int \delta\left[ k_z - \gamma(\mathbf{k}_\perp + \mathbf{k}'_{i\perp}) + \gamma(\mathbf{k}'_{i\perp}) \right] dk_z.$$

Using the property of the $\delta$ function, $\int \delta(k_z - a) dk_z = 1$, the integration over $k_z$ can be directly obtained, where the result takes the form

$$s = i\frac{\beta_0}{2n_0} \int d\boldsymbol{\varepsilon} \int d\mathbf{k}_\perp \chi(\boldsymbol{\varepsilon}) U_i^*(\boldsymbol{\varepsilon})$$
$$\int \left[ A_i(\mathbf{k}'_{i\perp}) A_d(\mathbf{k}'_{i\perp}) \right] A_d(\mathbf{k}_\perp + \mathbf{k}'_{i\perp}) e^{i\left[ (\mathbf{k}_\perp + \mathbf{k}'_{i\perp})\varepsilon_\perp + \gamma(\mathbf{k}_\perp + \mathbf{k}'_{i\perp})\varepsilon_z \right]} d\mathbf{k}'_{i\perp}. \quad (27)$$

Next, we see that the integration with respect to $\mathbf{k}'_{i\perp}$ can be expressed by correlation,

$$\int \left[ A_i(\mathbf{k}'_{i\perp}) A_d(\mathbf{k}'_{i\perp}) \right] A_d(\mathbf{k}_\perp + \mathbf{k}'_{i\perp}) e^{i\left[ (\mathbf{k}_\perp + \mathbf{k}'_{i\perp})\varepsilon_\perp + \gamma(\mathbf{k}_\perp + \mathbf{k}'_{i\perp})\varepsilon_z \right]} d\mathbf{k}'_{i\perp} = \left[ A_m(\mathbf{k}_\perp) \right] \otimes_{\mathbf{k}_\perp} \left[ A_d(\mathbf{k}_\perp) e^{i\left[ \mathbf{k}_\perp \varepsilon_\perp + \gamma(\mathbf{k}_\perp)\varepsilon_z \right]} \right], \quad (28)$$

$$A_m(\mathbf{k}_\perp) = \begin{cases} 1 & |\mathbf{k}_\perp| \le \min(\beta_0 NA_o, \beta_0 NA_c) \\ 0 & \text{else} \end{cases} \quad (29)$$

where $\otimes_{\mathbf{k}_\perp}$ indicates the correlation operation with respect to $\mathbf{k}_\perp$. In addition, because $A_m(\mathbf{k}_\perp)$ is real and centrosymmetric, the correlation operation is equivalent to the result of convolution. Applying the relations of Eq. (28) and (29), Eq. (27) can be written as

$$s = i\frac{\beta_0}{2n_0} \int d\boldsymbol{\varepsilon} \chi(\boldsymbol{\varepsilon}) U_i^*(\boldsymbol{\varepsilon})$$
$$\int \left[ A_m(\mathbf{k}_\perp) \right] \circledv_{\mathbf{k}_\perp} \left[ A_d(\mathbf{k}_\perp) e^{i\left[ \mathbf{k}_\perp \varepsilon_\perp + \gamma(\mathbf{k}_\perp)\varepsilon_z \right]} \right] d\mathbf{k}_\perp. \quad (30)$$

In addition, using the property Fourier transform, $\int f(k_x) dk_x = f(x)\big|_{x=0}$, the result of Eq. (30) can further be simplified

$$s = i\frac{\beta_0}{2n_0}\int d\boldsymbol{\varepsilon}\chi(\boldsymbol{\varepsilon})U_i^*(\boldsymbol{\varepsilon})\pi k_{min}^2 \text{Jinc}(\pi k_{min}\mathbf{r}_\perp)U_d(\mathbf{r}_\perp+\boldsymbol{\varepsilon}_\perp,\varepsilon_z)\Big|_{\mathbf{r}_\perp=0}$$
$$= i\frac{\beta_0}{4n_0}\pi k_{min}^2 \int \chi(\boldsymbol{\varepsilon})U_i^*(\boldsymbol{\varepsilon})U_d(\boldsymbol{\varepsilon})d\boldsymbol{\varepsilon}, \quad (31)$$

where $k_{min} = \min(\beta_0 NA_o, \beta_0 NA_c)$, and

$$U_d(\boldsymbol{\varepsilon}) = \int A_d(\mathbf{p})e^{i[\mathbf{p}\cdot\boldsymbol{\varepsilon}_\perp + \gamma(\mathbf{p})\varepsilon_z]}d\mathbf{p}. \quad (32)$$

$\text{Jinc}(\pi k_{min}\mathbf{r}_\perp)$ is the Fourier transform of a circular disc, and it has a value of 1/2 when $\pi k_{min}\mathbf{r}_\perp$ approaches zero. Eq. (31) suggests that the illumination beam interacts with the scattering potential, and only the scattered light within the detection aperture can be recorded by the photodetector. In practice, the illumination beam is scanned in 3D, while the specimen maintains a fixed position. This operation is equivalent to translating the specimen while keeping the illumination intact. Let $\boldsymbol{\rho}$ be the scanning vector (Fig. S3), such that Eq. (31) gives the final form of the measured signal as

$$s(\boldsymbol{\rho}) = i\frac{\beta_0}{4n_0}\pi k_{min}^2 \int \chi(\boldsymbol{\varepsilon}-\boldsymbol{\rho})U_d(\boldsymbol{\varepsilon})U_i^*(\boldsymbol{\varepsilon})d\boldsymbol{\varepsilon}$$
$$= i\frac{\beta_0}{4n_0}\pi k_{min}^2 \chi(\boldsymbol{\rho}) \circledast [U_d(\boldsymbol{\rho})U_i^*(\boldsymbol{\rho})]. \quad (33)$$

The simple expression in Eq. (33) shows the interferometric signal detected by the system is expressed as the output of a linear system, whereby the scattering potential input $\chi$, is convolved with the point spread function (PSF), $U_d(\boldsymbol{\rho})U_i^*(\boldsymbol{\rho})$. This PSF, particular to a laser-scanning system, represents the overlap (product) between the conjugated incident field and the detection function. The detection function $U_d$ can be pictured as the field obtained by focusing a plane wave through the detection optics, from the detector to the sample plane. In our case, $U_i^*$ is dominant, i.e., much narrower than $U_d$, as the illumination objective has a much higher NA than the detection one.

**Experimental and theoretical PSF**

The system's transfer function was estimated using a series of through-focus images of a micropillar. This approach of imaging a sharp edge to extract the system response has been used successfully in quantitative phase imaging, as the gradient of an edge image approximates well the PSF across that edge [33]. Figure S4 presents the phase images of a micropillar in LS-GLIM with different detection apertures. Z-stack of the dashed-box region was cropped to estimate the transfer function by differentiation along the y-axis. The comparison of experimental and theoretical transfer functions for LS-GLIM at varying condenser $NA_C$ openings is illustrated in Fig. S5. The objective used in the measurement was $40\times/1.3$. The z-sampling was chosen to be $0.25\ \mu m$. For visibility, the transfer functions are shown on a log plot. The experimental transfer functions show comparable frequency support with the theoretical results derived in Supplementary Note 1.

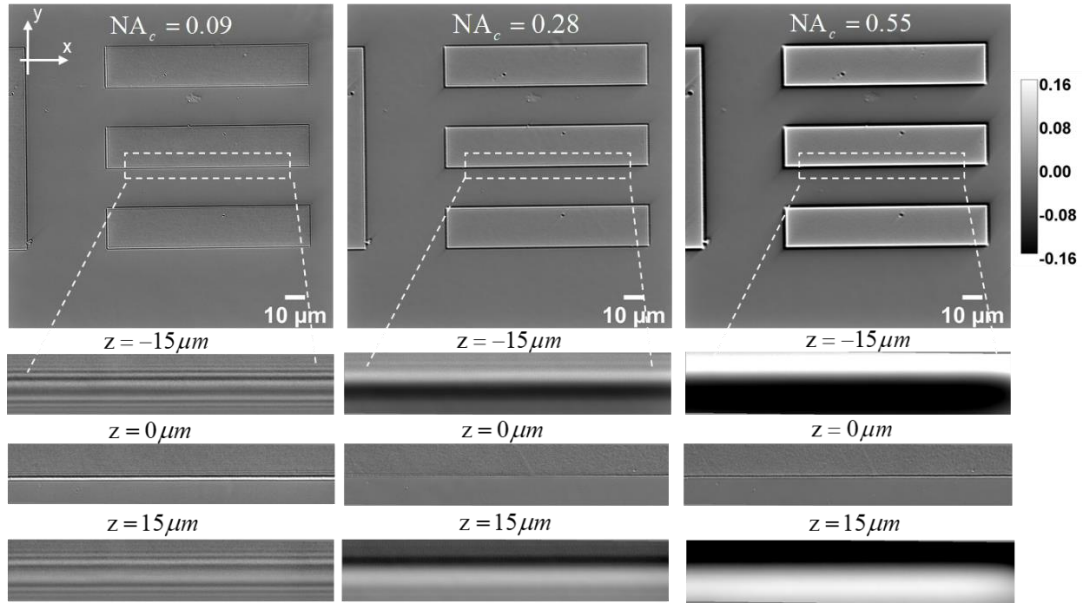

Fig. S4. Experimental phase edge measurement to estimate the transfer function in laser-scanning interference microscopy with different detection apertures.

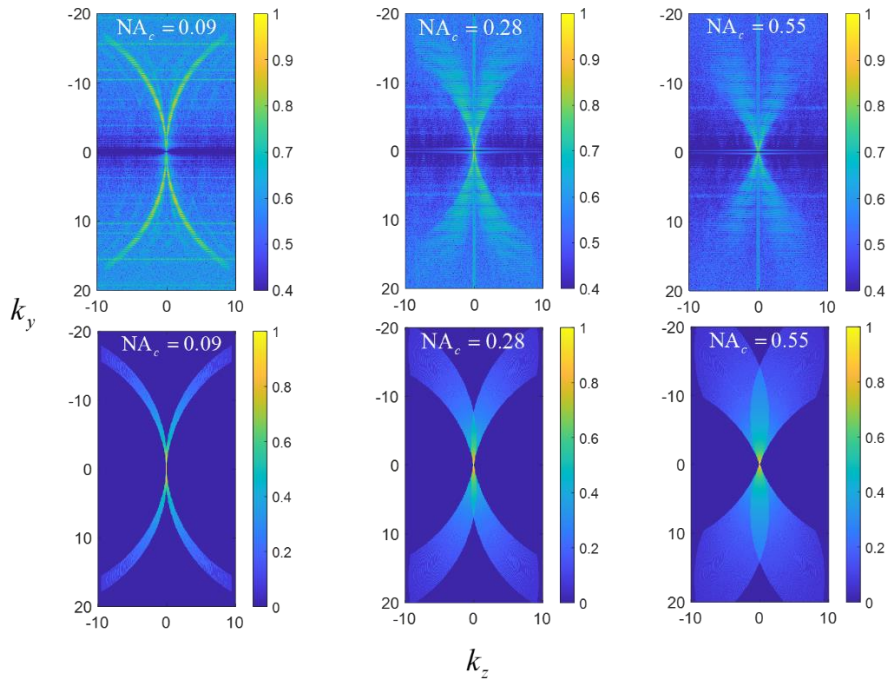

Fig. S5. Experimental (top) and theoretical (bottom) transfer functions of laser-scanning interference microscopy with different detection NAs.

**Supplementary Note 4: Network architectures and performances**

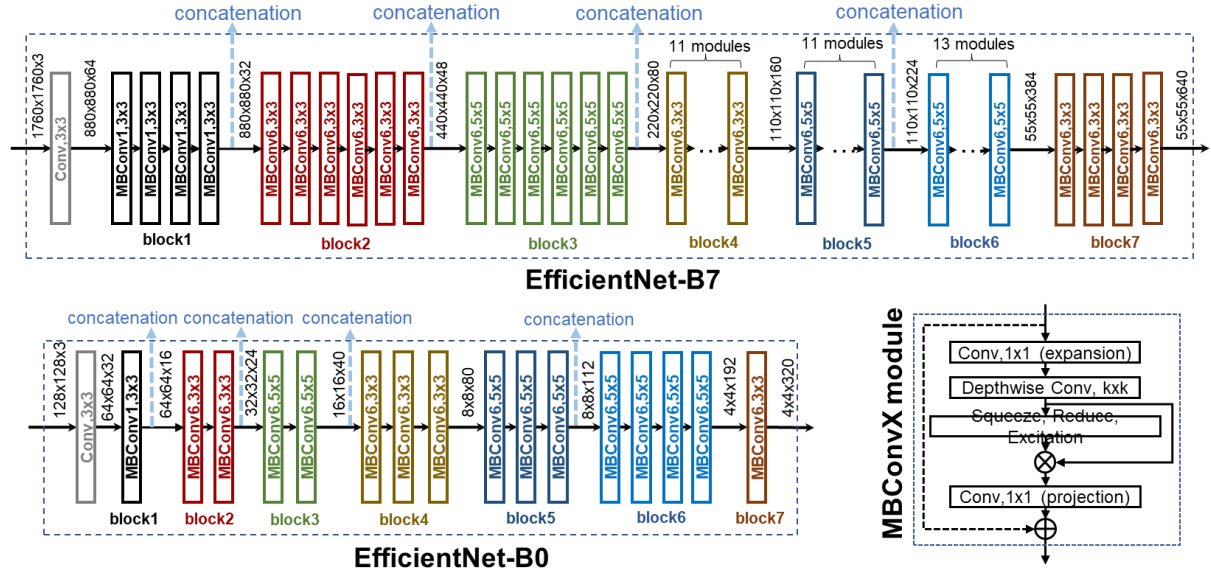

Fig. S6. The network architectures of EfficientNet-B7 and EfficientNet-B0 used in this study. The basic module in an EfficientNet is shown as MBConvX module. MBConv1 and MBConv6 indicate the ReLU and ReLU6 are used in the MBConvX module, respectively. The skip connection between the input and output of the MBConv6 module is not used in the first MBConvX module in each layer block. The feature maps for the four concatenation connections in E-U-Net are the outputs of block1, block2, block4, and block5, respectively, and are indicated as the dash blue arrows in Fig. 2a.

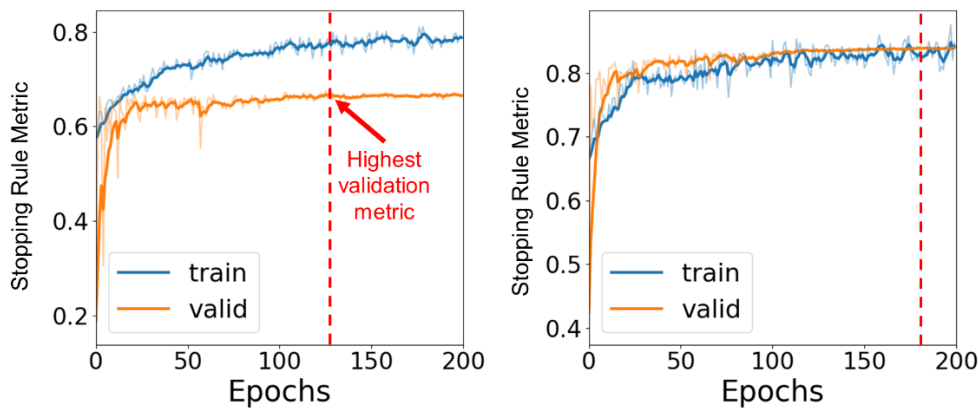

Fig. S7. Training and validation stopping rule metric curves over epochs related to training E-U-nets on neuron dataset in one of the three training procedures in the 3-fold cross-validation process. The left and right figures correspond to the network training for the fluorescent channels of Tau and MAP2 proteins, respectively. The red dash lines indicate the epoch when the network yielded the highest validation metric and was early stopped. The curves were smoothed for visualization by use of exponential moving average.

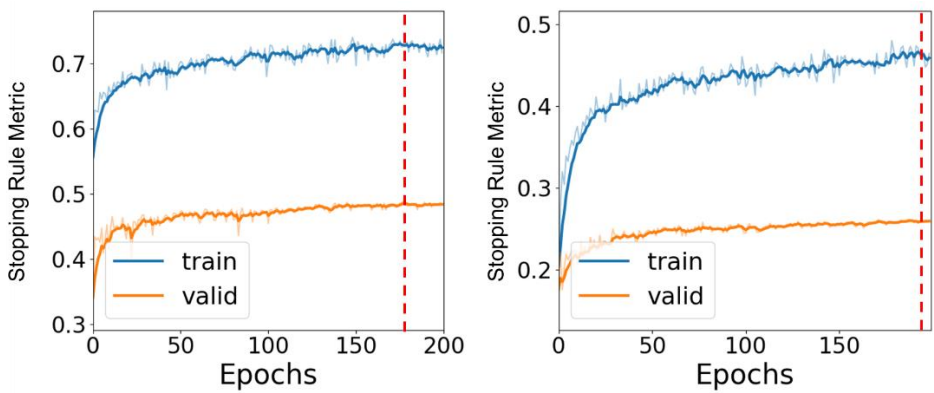

Fig. S8. Training and validation stopping rule metric over epochs related to training E-U-nets on spheroid cell dataset in one of the three procedures in the 3-fold cross-validation process. The left and right figures correspond to the network training for the two fluorescent channels of RNA and DNA, respectively. The red dash lines indicate the epoch when the network yielded the highest validation metric and was early stopped. The curves were smoothed for visualization by use of exponential moving average.

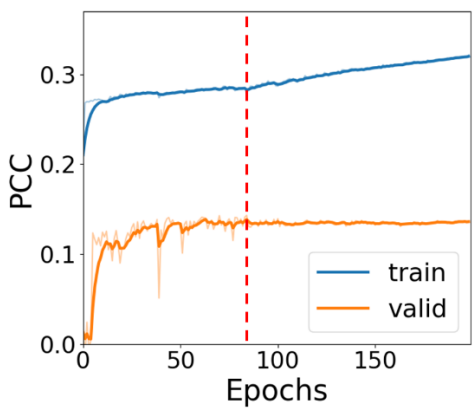

Fig. S9. Training and validation PCC scores over training epochs related to training E-U-Nets on bead dataset in one of the three procedures in the 3-fold cross-validation process. The red dash lines indicate the epoch when the network yielded the highest validation PCC and was early stopped.

The Pearson correlation and PSNR between the predicted and actual signals for microbeads, neurons, and spheroids are summarized in Figs. S10-S13. As typical of all fluorescence images, our ground truth measurements contained "salt-and-pepper" noise. To avoid biasing our comparison with irrelevant fluctuations in the signal-free background, we apply a 2D bilateral filter on the ground truth and the inference images of each z. The values of Pearson correlation and PSNR presented in Fig. S10-S13 are averaged values through each z-stack.

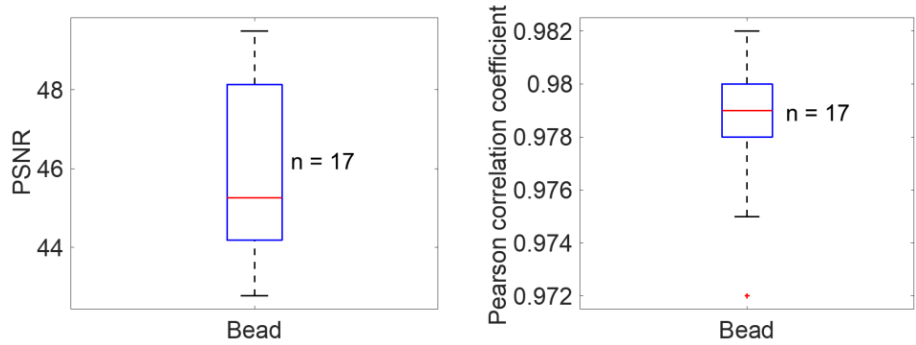

Fig. S10. 3-fold cross validation performances on the bead dataset.

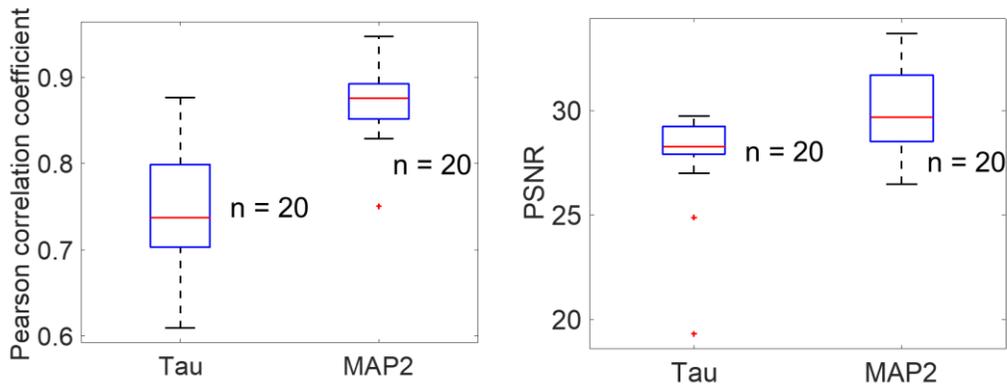

Fig. S11. 3-fold cross validation performance evaluated on the neuron dataset.

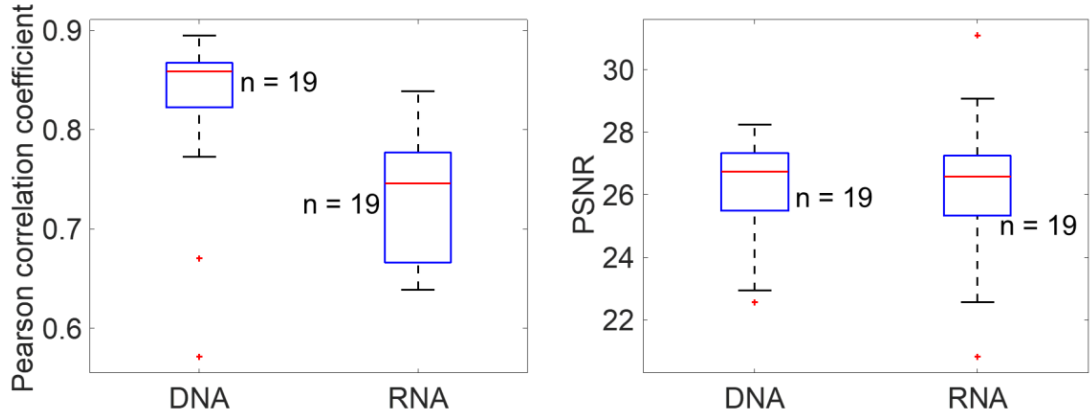

Fig. S12. 3-fold cross validation performance evaluated on spheroids dataset.

| Samples | channels | Model 1 | Model 2 | Model 3 |
|---|---|---|---|---|
| neuron_tau_map2_trial23_output | FL1 | 0.7923 | 0.8023 | 0.7893 |
|  | FL2 | 0.8978 | 0.9187 | 0.9149 |
| neuron_tau_map2_trial25_output | FL1 | 0.7286 | 0.7733 | 0.7150 |
|  | FL2 | 0.8942 | 0.9127 | 0.9059 |

| Samples | channels | Model 1 | Model 2 | Model 3 |
|---|---|---|---|---|
| neuron_tau_map2_trial23_output | FL1 | 26.71 | 27.17 | 26.68 |
|  | FL2 | 28.44 | 29.62 | 29.37 |
| neuron_tau_map2_trial25_output | FL1 | 29.09 | 29.95 | 28.99 |
|  | FL2 | 32.67 | 32.88 | 32.81 |

| Samples | channels | Model 1 | Model 2 | Model 3 |
|---|---|---|---|---|
| spheroids4_hydrogel_livedead_RNA_trial1_40x_output | FL1 | 0.8360 | 0.8340 | 0.8316 |
|  | FL2 | 0.6890 | 0.6957 | 0.6847 |
| spheroids4_livedead_RNA_trial7_40x_output | FL1 | 0.5988 | 0.5640 | 0.3182 |
|  | FL2 | 0.3654 | 0.4046 | 0.3902 |

| Samples | channels | Model 1 | Model 2 | Model 3 |
|---|---|---|---|---|
| spheroids4_hydrogel_livedead_RNA_trial1_40x_output | FL1 | 25.12 | 24.94 | 24.70 |
|  | FL2 | 28.26 | 30.34 | 28.24 |
| spheroids4_livedead_RNA_trial7_40x_output | FL1 | 23.91 | 23.23 | 22.65 |
|  | FL2 | 21.89 | 21.72 | 21.99 |

| Samples | Model 1 | Model 2 | Model 3 |
|---|---|---|---|
| bead-12 | 0.9779 | 0.9799 | 0.9796 |

| Samples | Model 1 | Model 2 | Model 3 |
|---|---|---|---|
| bead-12 | 44.0258 | 49.0219 | 45.8006 |

Fig. S13. The PCC and PSNR tested on unseen testing neuron, spheroid, and bead samples with the E-U-Nets trained in the 3-fold cross validation process. The left and right tables show the PCC and PSNR performances, respectively. The predicted fluorescent neuron and spheroid stacks, as well as their ground truth values were denoised by use of bilateral filters.

**Supplementary Video 1: 3D unstained live neuron prediction**

**Supplementary Video 2: Time-lapse unstained live neuron prediction**